\documentclass[a4paper,11pt]{article}
\usepackage[latin1]{inputenc}
\usepackage{amssymb}

\usepackage{times}

\title{{\bf   Infinite Games Specified by  2-Tape Automata}} 
\author{Olivier Finkel \\
{\it Equipe de Logique Math\'ematique}
\\Institut de Math\'ematiques de Jussieu
 \\  CNRS et Universit\'e Paris 7, France. \\ 
finkel@math.univ-paris-diderot.fr }

\date{}
\begin{document}

\newtheorem{theorem}{Theorem}[section]
\newtheorem{Rem}[theorem]{Remark}
\newtheorem{Exa}[theorem]{Example}

\newtheorem{Pro}[theorem]{Proposition}
\newtheorem{lem}[theorem]{Lemma}
\newtheorem{Cor}[theorem]{Corollary}
\newtheorem{defi}[theorem]{Definition}
\newtheorem{notation}[theorem]{Notation}

\def\ufootnote#1{\let\savedthfn\thefootnote\let\thefootnote\relax
\footnote{#1}\let\thefootnote\savedthfn\addtocounter{footnote}{-1}}

\newcommand{\bormxi}{{\bf\Pi}^{0}_{\xi}}
\newcommand{\bormlxi}{{\bf\Pi}^{0}_{<\xi}}
\newcommand{\bormz}{{\bf\Pi}^{0}_{0}}
\newcommand{\bormone}{{\bf\Pi}^{0}_{1}}
\newcommand{\ca}{{\bf\Pi}^{1}_{1}}
\newcommand{\bormtwo}{{\bf\Pi}^{0}_{2}}
\newcommand{\bormthree}{{\bf\Pi}^{0}_{3}}
\newcommand{\bormom}{{\bf\Pi}^{0}_{\omega}}
\newcommand{\borom}{{\bf\Delta}^{0}_{\omega}}
\newcommand{\borml}{{\bf\Pi}^{0}_{\lambda}}
\newcommand{\bormlpn}{{\bf\Pi}^{0}_{\lambda +n}}
\newcommand{\bormpm}{{\bf\Pi}^{0}_{1+m}}
\newcommand{\borapm}{{\bf\Sigma}^{0}_{1+m}}
\newcommand{\bormep}{{\bf\Pi}^{0}_{\eta +1}}
\newcommand{\borapxi}{{\bf\Sigma}^{0}_{\xi}}
\newcommand{\borai}{{\bf\Sigma}^{0}_{ 2.\xi +1 }}
\newcommand{\bormpxi}{{\bf\Pi}^{0}_{\xi}}
\newcommand{\bormpeta}{{\bf\Pi}^{0}_{1+\eta}}
\newcommand{\borapxipo}{{\bf\Sigma}^{0}_{\xi +1}}
\newcommand{\bormpxipo}{{\bf\Pi}^{0}_{\xi +1}}
\newcommand{\borpxi}{{\bf\Delta}^{0}_{\xi}}
\newcommand{\borel}{{\bf\Delta}^{1}_{1}}
\newcommand{\Borel}{{\it\Delta}^{1}_{1}}
\newcommand{\borone}{{\bf\Delta}^{0}_{1}}
\newcommand{\bortwo}{{\bf\Delta}^{0}_{2}}
\newcommand{\borthree}{{\bf\Delta}^{0}_{3}}
\newcommand{\boraone}{{\bf\Sigma}^{0}_{1}}
\newcommand{\boratwo}{{\bf\Sigma}^{0}_{2}}
\newcommand{\borathree}{{\bf\Sigma}^{0}_{3}}
\newcommand{\boraom}{{\bf\Sigma}^{0}_{\omega}}
\newcommand{\boraxi}{{\bf\Sigma}^{0}_{\xi}}
\newcommand{\ana}{{\bf\Sigma}^{1}_{1}}
\newcommand{\pca}{{\bf\Sigma}^{1}_{2}}
\newcommand{\Ana}{{\it\Sigma}^{1}_{1}}
\newcommand{\Boraone}{{\it\Sigma}^{0}_{1}}
\newcommand{\Borone}{{\it\Delta}^{0}_{1}}
\newcommand{\Bormone}{{\it\Pi}^{0}_{1}}
\newcommand{\Bormtwo}{{\it\Pi}^{0}_{2}}
\newcommand{\Ca}{{\it\Pi}^{1}_{1}}
\newcommand{\bormn}{{\bf\Pi}^{0}_{n}}
\newcommand{\bormm}{{\bf\Pi}^{0}_{m}}
\newcommand{\boralp}{{\bf\Sigma}^{0}_{\lambda +1}}
\newcommand{\borat}{{\bf\Sigma}^{0}_{|\theta |}}
\newcommand{\bormat}{{\bf\Pi}^{0}_{|\theta |}}
\newcommand{\Borapxi}{{\it\Sigma}^{0}_{\xi}}
\newcommand{\Bormpxipo}{{\it\Pi}^{0}_{1+\xi +1}}
\newcommand{\Borapn}{{\it\Sigma}^{0}_{1+n}}
\newcommand{\borapn}{{\bf\Sigma}^{0}_{1+n}}
\newcommand{\boraxipm}{{\bf\Sigma}^{0}_{\xi^\pm}}
\newcommand{\Boratwo}{{\it\Sigma}^{0}_{2}}
\newcommand{\Borathree}{{\it\Sigma}^{0}_{3}}
\newcommand{\Borapnpo}{{\it\Sigma}^{0}_{1+n+1}}
\newcommand{\Bormpxi}{{\it\Pi}^{0}_{\xi}}
\newcommand{\Borpxi}{{\it\Delta}^{0}_{\xi}}
\newcommand{\boratpxi}{{\bf\Sigma}^{0}_{2+\xi}}
\newcommand{\Boratpxi}{{\it\Sigma}^{0}_{2+\xi}}
\newcommand{\bormltpxi}{{\bf\Pi}^{0}_{<2+\xi}}
\newcommand{\Bormltpxi}{{\it\Pi}^{0}_{<2+\xi}}
\newcommand{\borapeap}{{\bf\Sigma}^{0}_{1+\eta_{\alpha ,p}}}
\newcommand{\borapeapn}{{\bf\Sigma}^{0}_{1+\eta_{\alpha ,p,n}}}
\newcommand{\Borapeap}{{\it\Sigma}^{0}_{1+\eta_{\alpha ,p}}}
\newcommand{\Bormpn}{{\it\Pi}^{0}_{1+n}}
\newcommand{\Borpn}{{\it\Delta}^{0}_{1+n}}
\newcommand{\borapximo}{{\bf\Sigma}^{0}_{1+(\xi -1)}}
\newcommand{\borpeta}{{\bf\Delta}^{0}_{1+\eta}}

\newcommand{\hs}{\hspace{12mm}

}
\newcommand{\noi}{\noindent}

\newcommand{\om}{\omega}
\newcommand{\Si}{\Sigma}
\newcommand{\Sis}{\Sigma^\star}
\newcommand{\Sio}{\Sigma^\omega}
\newcommand{\nl}{\newline}
\newcommand{\lra}{\leftrightarrow}
\newcommand{\fa}{\forall}
\newcommand{\ra}{\rightarrow}
\newcommand{\orl}{ $\omega$-regular language}

\newcommand{\Ga}{\Gamma}
\newcommand{\Gas}{\Gamma^\star}
\newcommand{\Gao}{\Gamma^\omega}
\newcommand{\ite}{\item}
\newcommand{\la}{language}
\newcommand{\Lp}{L(\varphi)}
\newcommand{\abs}{\{a, b\}^\star}
\newcommand{\abcs}{\{a, b, c \}^\star}
\newcommand{\ol}{$\omega$-language}

\newcommand{\tla}{\twoheadleftarrow}
\newcommand{\de}{deterministic }
\newcommand{\proo}{\noi {\bf Proof.} }
\newcommand {\ep}{\hfill $\square$}

\maketitle

\begin{abstract}
\noi We prove that  the determinacy of   Gale-Stewart games whose winning sets are infinitary rational relations accepted by 
  $2$-tape B\"uchi automata  is  equivalent to the determinacy of (effective) analytic Gale-Stewart games which is known to be a large cardinal 
assumption.  
Then we prove that winning strategies, when they exist, 
 can be very complex,  i.e. highly non-effective,   in  these games. We  prove the same results  for  
  Gale-Stewart games with winning sets accepted by real-time $1$-counter  B\"uchi   automata,  then 
extending previous results obtained about these games.

\begin{enumerate}
\item
  There exists a $2$-tape B\"uchi automaton (respectively, a   real-time $1$-counter B\"uchi automaton)
$\mathcal{A}$ such that: (a) there is a model  of ZFC in which  
Player 1 has a winning strategy $\sigma$ in the  
game $G(L(\mathcal{A}))$ but $\sigma$ cannot be recursive and not even in the class $(\Sigma_2^1 \cup \Pi_2^1)$;  
(b) there is a model  of ZFC in which the game  $G(L(\mathcal{A}))$ 
is not determined. 
\item   There exists a $2$-tape B\"uchi automaton (respectively, a   real-time $1$-counter B\"uchi automaton) 
$\mathcal{A}$ such that  $L(\mathcal{A})$ is an arithmetical 
$\Delta_3^0$-set and  Player 2 has a winning strategy in the game $G(L(\mathcal{A}))$ 
 but has no hyperarithmetical winning strategies in this game. 
\item There exists a recursive sequence of  $2$-tape  B\"uchi automata (respectively, of  real-time $1$-counter B\"uchi automata) $\mathcal{A}_n$,  
$n\geq 1$,  such that all games  $G(L(\mathcal{A}_n))$ are determined, but for which it is $\Pi_2^1$-complete hence 
highly undecidable to determine whether Player 1 has a winning strategy in the game $G(L(\mathcal{A}_n))$.  
\end{enumerate}

\noi Then we consider  the strenghs of determinacy for these games, and we prove the following results.  
\begin{enumerate}
\item There exists  a 2-tape B\"uchi   automaton (respectively, a   real-time $1$-counter B\"uchi automaton) $A_\sharp$ 
such  that the game $G(A_\sharp)$   is determined iff the effective analytic determinacy 
holds. 
\item There is a  transfinite sequence of 2-tape B\"uchi   automata (respectively, of  real-time $1$-counter B\"uchi automata)  ($\mathcal{A}_\alpha$)$_{\alpha<\om_1^{\rm{CK}}}$, 
indexed by recursive ordinals, such that the games $G(L(\mathcal{A}_\alpha))$ have strictly  increasing strenghs of determinacy. 
\end{enumerate}

\hs \noi We show also that the determinacy of Wadge games between two players in charge of infinitary rational relations accepted by 
  $2$-tape B\"uchi automata
  is  equivalent to the  (effective) analytic  Wadge determinacy and thus also equivalent to the (effective) analytic  determinacy. 
\end{abstract}


\section{Introduction}

In Computer Science,  non terminating systems in relation with an environment may be specified  with some particular infinite games of perfect information, 
called  Gale Stewart games since they have been firstly studied by Gale and Stewart in 1953  in \cite{Gale-Stewart53}. 
The two players in such a game are respectively  a non terminating reactive
 program and  the  ``environment".  A Gale-Stewart game is defined as follows. 
If $X$ is a (countable) alphabet having at 
least two letters and $A \subseteq X^\om$, then the Gale-Stewart game $G(A)$ is an infinite  game with perfect
 information between two players. Player 1 first writes a letter
$a_1\in X$, then Player 2 writes a letter $b_1\in X$,
 then Player 1 writes $a_2\in X$, and so on $\ldots$
After $\om$ steps, the two players have composed an infinite word 
$x =a_1b_1a_2b_2\ldots$ of $X^\om$.
 Player 1 wins the play iff $x \in A$, otherwise Player 2
wins the play. The game $G(A)$ is said to be determined iff 
one of the two players has a winning strategy.

 Then the problem of the synthesis of winning strategies
is of great practical
interest for the problem of program synthesis in reactive systems.
In particular, if $A \subseteq X^\om$, where $X$ is here a finite alphabet, and $A$ is effectively presented, i.e. accepted by a given 
finite machine or defined by a given logical formula, the following questions naturally arise, see  \cite{Thomas95,LescowThomas}:
(1) ~ Is the game $G(A)$ determined?
~ (2) ~ If Player 1 has a winning strategy, is it effective, i.e. computable? 
~(3) ~ What are the amounts of space and time necessary to compute such a winning strategy? 
B\"uchi and Landweber  gave a solution to the famous Church's Problem, posed in 1957,   by proving  that in a Gale Stewart game $G(A)$,
where $A$ is a regular $\om$-language, one can decide who the winner  is  and
compute a winning strategy given by a finite state transducer, see \cite{Thomas08}. 
 Walukiewicz  extended B\"uchi and Landweber's Theorem to the case of a 
winning set  $A$  which is deterministic context-free, i.e. accepted by some deterministic pushdown automaton, answering a question of 
Thomas and Lescow in \cite{Thomas95,LescowThomas}. He first showed  in  \cite{wal} 
that one can effectively construct winning strategies in parity games played on
pushdown graphs and that these strategies can be computed by pushdown
transducers. 
Notice that later some extensions to the case of higher-order pushdown automata have been established  \cite{Cachat03,CHMOS08}. 

In \cite{Fin12,Fin13-JSL} we have studied Gale-Stewart games $G(A)$, 
where $A$ is a context-free $\om$-language accepted by a {\it non-deterministic}  pushdown automaton, or 
even by a $1$-counter automaton.  We have proved that the determinacy of   
Gale-Stewart games  $G(A)$, whose winning sets $A$ are accepted by 
real-time  $1$-counter B\"uchi automata,  is  equivalent to the determinacy of (effective) analytic Gale-Stewart games. 
On the other hand  Gale-Stewart  games have been much studied in Set Theory and  in Descriptive Set Theory, see \cite{Kechris94,Jech}. 
It has been proved by Martin that every Gale-Stewart game
 $G(A)$, where $A$ is a Borel set, is determined \cite{Kechris94}. Notice that this is proved in ZFC, the commonly accepted axiomatic 
framework for Set Theory in which all usual mathematics can be developped.  But  the determinacy of Gale-Stewart games $G(A)$,  where 
$A$ is an (effective) analytic set, is not provable in ZFC; Martin and Harrington have proved that it is  a large cardinal 
assumption equivalent to the existence of a particular real, called the real $0^\sharp$, see \cite[page  637]{Jech}. 
Thus we proved in \cite{Fin12,Fin13-JSL} that the determinacy of Gale-Stewart games  $G(A)$, whose winning sets $A$ are accepted by 
real-time  $1$-counter B\"uchi automata, is also 
 equivalent to the existence of the real $0^\sharp$, and thus not provable in ZFC. 

In this paper we consider Gale-Stewart games $G(L(\mathcal{A}))$,  where $L(\mathcal{A})$ is  an infinitary rational relation, i.e. an $\om$-language over a product alphabet
$X=\Si \times \Gamma$, which is accepted by a $2$-tape (non-deterministic) B\"uchi automaton $\mathcal{A}$.  
In such a game, the two players alternatively write letters from the product alphabet $X=\Si \times \Gamma$, and after $\om$ steps they have produced 
an infinite word over $X$ which may be identified with a pair of infinite words $(u,v) \in \Sio \times \Ga^\om$. 
Then Player 1 wins the play if  $(u,v) \in L(\mathcal{A})$. Notice that if the $2$-tape  B\"uchi automaton $\mathcal{A}$ is synchronous then the 
winning set is actually a regular $\om$-language over the product alphabet  $X=\Si \times \Gamma$. 
Then the infinitary rational relation $L(\mathcal{A})$ is Borel, the game  $G(L(\mathcal{A}))$ is determined,  and      it follows 
from B\"uchi and Landweber's Theorem that one can decide who the winner  is  and
compute a winning strategy given by a finite state transducer.  We show in this paper that the situation is very different when the $2$-tape  B\"uchi  automaton 
may be asynchronous. 

We firstly prove that  the determinacy of   Gale-Stewart games 
whose winning sets are infinitary rational relations accepted by 
  $2$-tape B\"uchi automata  is  equivalent to the determinacy of  Gale-Stewart games whose  winning sets are   accepted by 
  $1$-counter  B\"uchi automata   and thus also 
 equivalent to the existence of the real $0^\sharp$.  In particular, it is not provable in ZFC. 

Next  we prove numerous more results on these games along with similar results about $1$-counter games which extend the previous results 
obtained in \cite{Fin12,Fin13-JSL}. In particular,  we prove that 
winning strategies in these games, when they exist, 
 can be very complex,  i.e. highly non-effective. 

\begin{enumerate}

\item
  There exists a $2$-tape B\"uchi automaton (respectively, a  real-time  $1$-counter B\"uchi automaton)   
$\mathcal{A}$ such that: (a) there is a model  of ZFC in which  
Player 1 has a winning strategy $\sigma$ in the  
game $G(L(\mathcal{A}))$ but $\sigma$ cannot be recursive and not even in the class $(\Sigma_2^1 \cup \Pi_2^1)$;  
(b) there is a model  of ZFC in which the game  $G(L(\mathcal{A}))$ 
is not determined. 

\item   There exists a $2$-tape B\"uchi automaton     (respectively, a   real-time $1$-counter B\"uchi automaton)     
 $\mathcal{A}$ such that the infinitary rational relation    (respectively,  the $1$-counter $\om$-language) $L(\mathcal{A})$ is an arithmetical 
$\Delta_3^0$-set and  Player 2 has a winning strategy in the game $G(L(\mathcal{A}))$ 
 but has no hyperarithmetical winning strategies in this game. 

\item There exists a recursive sequence of  $2$-tape  B\"uchi automata (respectively, of real-time   $1$-counter B\"uchi automata)   $\mathcal{A}_n$,  
$n\geq 1$,  such that all games  $G(L(\mathcal{A}_n))$ are determined, but for which it is $\Pi_2^1$-complete, hence 
highly undecidable,  to determine whether Player 1 has a winning strategy in the game $G(L(\mathcal{A}_n))$.  

\end{enumerate}

\noi Then we consider  the possible strenghs of determinacy for these games, and prove the following results, using results of  Harrington and Stern 
on effective analytic games, \cite{Harrington,Stern82}.  

\begin{enumerate}
\item There exists  a 2-tape B\"uchi   automaton (respectively, a  real-time  $1$-counter B\"uchi automaton)      $A_\sharp$ 
such  that the game $G(L(A_\sharp))$   is determined iff the effective analytic determinacy 
holds. 
\item There is a  transfinite sequence of 2-tape B\"uchi   automata  (respectively, of  real-time   $1$-counter B\"uchi automata) 
  ($\mathcal{A}_\alpha$)$_{\alpha<\om_1^{\rm{CK}}}$, 
indexed by recursive ordinals, such that the games $G(L(\mathcal{A}_\alpha))$ have strictly  increasing strenghs of determinacy. 
\end{enumerate}

On the other hand, there is another class of infinite games of perfect information which have been much studied in Set Theory 
and  in Descriptive Set Theory: the Wadge games firstly studied  by  Wadge     in \cite{Wadge83} where he determined 
a great refinement of the Borel hierarchy defined 
via  the notion of  reduction by continuous functions. 
  The Wadge games are closely related to the notion of reducibility by continuous  functions.  
For $L\subseteq X^\om$ and $L'\subseteq Y^\om$, $L$ is said to be Wadge reducible to $L'$ 
 iff there exists a continuous function $f: X^\om \ra Y^\om$, such that
$L=f^{-1}(L')$; this is then denoted by $L\leq _W L'$.  
On the other hand, the Wadge game $W(L, L')$ is  an infinite  game with perfect information between two players,
Player 1 who is in charge of $L$ and Player 2 who is in charge of $L'$.  And it turned out that  Player 2 has a winning strategy in the Wadge game 
 $W(L, L')$  iff  $L\leq _W L'$. 
The Wadge games have also been considered in Computer Science since they are important in the study of the topological complexity 
of languages of infinite words or trees accepted by various kinds of automata, 
\cite{PerrinPin,Staiger97,Fin-mscs06,Fink-Wd,Selivanov03b,Selivanov08,ADNM}. 
We proved in \cite{Fin12,Fin13-JSL} that the determinacy of Wadge games between two players in charge of 
$\om$-languages accepted by  real-time $1$-counter B\"uchi automata  is  equivalent to the  (effective) analytic  Wadge determinacy, which is known to be 
equivalent to the    (effective) analytic  determinacy  (see \cite{Louveau-Saint-Raymond})   and thus also 
equivalent to the existence of the real $0^\sharp$. 
We consider here Wadge games between two players in charge of 
infinitary rational relations accepted by  $2$-tape B\"uchi automata and we prove that the determinacy of these Wadge games is equivalent to 
the determinacy of Wadge games between two players in charge of 
$\om$-languages accepted by  real-time $1$-counter B\"uchi automata and thus also equivalent to the (effective) analytic determinacy.  In 
particular,  the determinacy of these games is not provable in ZFC.

 Notice that as  the results presented in this paper might be of interest to  both set theorists and theoretical computer scientists, we 
shall recall in detail some notions of automata theory which are well known to computer scientists but not to set theorists. In a similar 
way,  we give  a presentation of  some  results of set theory which are well known to  set theorists   but not to   computer scientists.

The paper is organized as follows. We recall some known notions in Section 2. We study Gale-Stewart games with winning sets accepted by 
2-tape B\"uchi   automata  or by   $1$-counter B\"uchi automata in Section 3. 
In Section 4 we study Wadge  games between two players in charge of infinitary rational  relations. Some concluding remarks are given in Section 5.

\section{Recall of some known notions}
 
~~~~~  We assume   the reader to be familiar with the theory of formal ($\om$-)languages  
\cite{Staiger97,PerrinPin}.
We recall the  usual notations of formal language theory. 

If  $\Si$ is a finite or countably infnite alphabet, a {\it non-empty finite word} over $\Si$ is any 
sequence $x=a_1\ldots a_k$, where $a_i\in\Sigma$ 
for $i=1,\ldots ,k$ , and  $k$ is an integer $\geq 1$. The {\it length}
 of $x$ is $k$, denoted by $|x|$.
 The {\it empty word}  is denoted by $\lambda$; its length is $0$. 
 $\Sis$  is the {\it set of finite words} (including the empty word) over $\Sigma$.
A  (finitary) {\it language} $V$ over an alphabet $\Sigma$ is a subset of  $\Sis$.

 The {\it first infinite ordinal} is $\om$.
 An $\om$-{\it word} over $\Si$ is an $\om$ -sequence $a_1 \ldots a_n \ldots$, where for all 
integers $ i\geq 1$, ~
$a_i \in\Sigma$.  When $\sigma=a_1 \ldots a_n \ldots$ is an $\om$-word over $\Si$, we write
 $\sigma(n)=a_n$,   $\sigma[n]=\sigma(1)\sigma(2)\ldots \sigma(n)$  for all $n\geq 1$ and $\sigma[0]=\lambda$.

 The usual concatenation product of two finite words $u$ and $v$ is 
denoted $u.v$ (and sometimes just $uv$). This product is extended to the product of a 
finite word $u$ and an $\om$-word $v$: the infinite word $u.v$ is then the $\om$-word such that:

 $(u.v)(k)=u(k)$  if $k\leq |u|$ , and 
 $(u.v)(k)=v(k-|u|)$  if $k>|u|$.
  
 The {\it set of } $\om$-{\it words} over  the alphabet $\Si$ is denoted by $\Si^\om$.
An  $\om$-{\it language} $V$ over an alphabet $\Sigma$ is a subset of  $\Si^\om$, and its  complement (in $\Sio$) 
 is $\Sio - V$, denoted $V^-$.

  The {\it prefix relation} is denoted $\sqsubseteq$: a finite word $u$ is a {\it prefix} 
of a finite word $v$ (respectively,  an infinite word $v$), denoted $u\sqsubseteq v$,  
 if and only if there exists a finite word $w$ 
(respectively,  an infinite word $w$), such that $v=u.w$.  

If  $L$ is a finitary language (respectively,  an $\om$-language) over   the alphabet $\Si$ then the  set 
${\rm Pref}(L)$ of prefixes of elements of $L$ is defined by  ${\rm Pref}(L)=\{ u\in \Sis \mid  \exists v\in L ~~ u \sqsubseteq v \}$.

 We now recall the definition of $k$-counter B\"uchi automata, reading $\om$-words over a {\it finite} alphabet, 
 which will be useful in the sequel. 

 Let $k$ be an integer $\geq 1$. 
A  $k$-counter machine has $k$ {\it counters}, each of which containing a  non-negative integer. 
The machine can test whether the content of a given counter is zero or not. 
And transitions depend on the letter read by the machine, the current state of the finite control, and the tests about the values of the counters. 
Notice that in this model some  $\lambda$-transitions are allowed. During these transitions the reading head of the machine does not move to the right, i.e. 
 the machine does not  read any more letter. 

Formally a  $k$-counter machine is a 4-tuple 
$\mathcal{M}$=$(K,\Si,$ $ \Delta, q_0)$,  where $K$ 
is a finite set of states, $\Sigma$ is a finite input alphabet, 
 $q_0\in K$ is the initial state, 
and  $\Delta \subseteq K \times ( \Si \cup \{\lambda\} ) \times \{0, 1\}^k \times K \times \{0, 1, -1\}^k$ is the transition relation. 
The $k$-counter machine $\mathcal{M}$ is said to be {\it real time} iff: 
$\Delta \subseteq K \times
  \Si \times \{0, 1\}^k \times K \times \{0, 1, -1\}^k$, 
 i.e. iff there are no  $\lambda$-transitions. 

If  the machine $\mathcal{M}$ is in state $q$ and 
$c_i \in \mathbf{N}$ is the content of the $i^{th}$ counter 
 $\mathcal{C}$$_i$ then 
the  configuration (or global state)
 of $\mathcal{M}$ is the  $(k+1)$-tuple $(q, c_1, \ldots , c_k)$.

 For $a\in \Si \cup \{\lambda\}$, 
$q, q' \in K$ and $(c_1, \ldots , c_k) \in \mathbf{N}^k$ such 
that $c_j=0$ for $j\in E \subseteq  \{1, \ldots , k\}$ and $c_j >0$ for 
$j\notin E$, if 
$(q, a, i_1, \ldots , i_k, q', j_1, \ldots , j_k) \in \Delta$ where $i_j=0$ for $j\in E$ 
and $i_j=1$ for $j\notin E$, then we write:

~~~~~~~~$a: (q, c_1, \ldots , c_k)\mapsto_{\mathcal{M}} (q', c_1+j_1, \ldots , c_k+j_k)$.

 Thus  the transition relation must obviously satisfy:
 \nl if $(q, a, i_1, \ldots , i_k, q', j_1, \ldots , j_k)  \in    \Delta$ and  $i_m=0$ for 
 some $m\in \{1, \ldots , k\}$  then $j_m=0$ or $j_m=1$ (but $j_m$ may not be equal to $-1$). 
  
Let $\sigma =a_1a_2 \ldots a_n \ldots $ be an $\om$-word over $\Si$. 
An $\om$-sequence of configurations $r=(q_i, c_1^{i}, \ldots c_k^{i})_{i \geq 1}$ is called 
a run of $\mathcal{M}$ on $\sigma$  iff:

(1)  $(q_1, c_1^{1}, \ldots c_k^{1})=(q_0, 0, \ldots, 0)$

(2)   for each $i\geq 1$, there  exists $b_i \in \Si \cup \{\lambda\}$ such that
 $b_i: (q_i, c_1^{i}, \ldots c_k^{i})\mapsto_{\mathcal{M}}  
(q_{i+1},  c_1^{i+1}, \ldots c_k^{i+1})$  
and such that  ~  $a_1a_2\ldots a_n\ldots =b_1b_2\ldots b_n\ldots$

For every such run $r$, $\mathrm{In}(r)$ is the set of all states entered infinitely
 often during $r$.

\begin{defi} A B\"uchi $k$-counter automaton  is a 5-tuple 
$\mathcal{M}$=$(K,\Si,$ $\Delta, q_0, F)$, 
where $ \mathcal{M}'$=$(K,\Si,$  $\Delta, q_0)$
is a $k$-counter machine and $F \subseteq K$ 
is the set of accepting  states.
The \ol~ accepted by $\mathcal{M}$ is:~~ $L(\mathcal{M})$= $\{  \sigma\in\Si^\om \mid \mbox{  there exists a  run r
 of } \mathcal{M} \mbox{ on } \sigma \mbox{  such that } \mathrm{In}(r)
 \cap F \neq \emptyset \}$

\end{defi}

  The class of \ol s accepted by  B\"uchi $k$-counter automata  is  
denoted ${\bf BCL}(k)_\om$.
 The class of \ol s accepted by {\it  real time} B\"uchi $k$-counter automata  will be 
denoted {\bf r}-${\bf BCL}(k)_\om$.
  The class ${\bf BCL}(1)_\om$ is  a strict subclass of the class ${\bf CFL}_\om$ of context free \ol s
accepted by B\"uchi pushdown automata.

\hs 
Infinitary rational relations 
are subsets of  $\Sio \times \Gao$,  where 
$\Si$ and  $\Ga$ are finite alphabets, which are accepted by 
$2$-tape B\"uchi automata. 

\begin{defi}
A  2-tape B\"uchi automaton 
 is a sextuple $\mathcal{A}=(K, \Si, \Ga, \Delta, q_0, F)$, where 
$K$ is a finite set of states, $\Si$ and $\Ga$ are finite  alphabets, 
$\Delta$ is a finite subset of $K \times \Sis \times \Gas \times K$ called 
the set of transitions, $q_0$ is the initial state,  and $F \subseteq K$ is the set of 
accepting states. 
\nl A computation $\mathcal{C}$ of the  
2-tape B\"uchi automaton $\mathcal{A}$ is an infinite sequence of transitions 
$$(q_0, u_1, v_1, q_1), (q_1, u_2, v_2, q_2), \ldots ,(q_{i-1}, u_{i}, v_{i}, q_{i}), 
(q_i, u_{i+1}, v_{i+1}, q_{i+1}), \ldots $$
\noi The computation is said to be successful iff there exists a final state $q_f \in F$ 
and infinitely many integers $i\geq 0$ such that $q_i=q_f$. 
\nl The input word of the computation is $u=u_1.u_2.u_3 \ldots$
\nl The output word of the computation is $v=v_1.v_2.v_3 \ldots$
\nl Then the input and the output words may be finite or infinite. 
\nl The infinitary rational relation $L(\mathcal{A})\subseteq \Sio \times \Ga^\om$ 
accepted by the 2-tape B\"uchi automaton $\mathcal{A}$ 
is the set of pairs $(u, v) \in \Sio \times \Ga^\om$ such that $u$ and $v$ are the input 
and the output words of some successful computation $\mathcal{C}$ of $\mathcal{A}$. 
\nl The set of infinitary rational relations will be denoted by ${\bf RAT}_\om$. 
\end{defi}

\hs  We assume the reader to be familiar with basic notions of topology which
may be found in \cite{Kechris94,LescowThomas,Staiger97,PerrinPin}.
There is a natural metric on the set $\Sio$ of  infinite words 
over a finite or countably infinite alphabet 
$\Si$ containing at least two letters which is called the {\it prefix metric} and is defined as follows. For $u, v \in \Sio$ and 
$u\neq v$ let $\delta(u, v)=2^{-l_{\mathrm{pref}(u,v)}}$ where $l_{\mathrm{pref}(u,v)}$ 
 is the first integer $n$
such that the $(n+1)^{st}$ letter of $u$ is different from the $(n+1)^{st}$ letter of $v$. 
This metric induces on $\Sio$ the usual   topology in which the {\it open subsets} of 
$\Sio$ are of the form $W.\Si^\om$, for $W\subseteq \Sis$.  
A set $L\subseteq \Si^\om$ is a {\it closed set} iff its complement $\Si^\om - L$ 
is an open set. If  the alphabet $\Si$ is finite then the set $\Sio$ equipped with this topology is a Cantor space, and if $\Si=\om$ then 
the set $\om^\om$ equipped with this topology is the classical Baire space. We  shall consider only these two cases in the sequel. 

For $V \subseteq \Sis$ we denote ${\rm Lim}(V)=\{ x \in \Sio \mid  \exists^\infty  n \geq 1  ~~ x[n] \in V \}$ the set of infinite words over 
$\Si$ having infinitely many prefixes in $V$.  Then the topological closure ${\rm Cl}(L)$ of a set $L\subseteq \Si^\om$ is equal to 
${\rm Lim}({\rm Pref}(L))$. Thus we have also the following characterization of closed subsets of $\Si^\om$:  
a set  $L\subseteq \Si^\om$ is a closed subset of the  space $\Si^\om$ iff  
$L={\rm Lim}({\rm Pref}(L))$.

We   now recall 
the definition of the {\it Borel Hierarchy} of subsets of $X^\om$. 

\begin{defi}
For a non-null countable ordinal $\alpha$, the classes ${\bf \Si}^0_\alpha$
 and ${\bf \Pi}^0_\alpha$ of the Borel Hierarchy on the topological space $X^\om$ 
are defined as follows:
 ${\bf \Si}^0_1$ is the class of open subsets of $X^\om$, 
 ${\bf \Pi}^0_1$ is the class of closed subsets of $X^\om$, 
 and for any countable ordinal $\alpha \geq 2$: 
\nl ${\bf \Si}^0_\alpha$ is the class of countable unions of subsets of $X^\om$ in 
$\bigcup_{\gamma <\alpha}{\bf \Pi}^0_\gamma$.
 \nl ${\bf \Pi}^0_\alpha$ is the class of countable intersections of subsets of $X^\om$ in 
$\bigcup_{\gamma <\alpha}{\bf \Si}^0_\gamma$.

A set $L\subseteq X^\om$ is Borel iff it is in the union $\bigcup_{\alpha < \om_1} {\bf \Si}^0_\alpha = \bigcup_{\alpha < \om_1} {\bf \Pi}^0_\alpha$, where  
$\om_1$ is the first uncountable ordinal. 
\end{defi}

\noi    
There are also some subsets of $X^\om$ which are not Borel. 
In particular,  
the class of Borel subsets of $X^\om$ is strictly included into 
the class  ${\bf \Si}^1_1$ of {\it analytic sets} which are 
obtained by projection of Borel sets. The  {\it co-analytic sets}  are the complements of 
analytic sets.  

\begin{defi} 
A subset $A$ of  $X^\om$ is in the class ${\bf \Si}^1_1$ of {\it analytic} sets
iff there exist a finite alphabet $Y$ and a Borel subset $B$  of  $(X \times Y)^\om$ 
such that $ x \in A \lra \exists y \in Y^\om $ such that $(x, y) \in B$, 
where $(x, y)$ is the infinite word over the alphabet $X \times Y$ such that
$(x, y)(i)=(x(i),y(i))$ for each  integer $i\geq 1$.
\end{defi} 

We now recall the notion of  completeness with regard to reduction by continuous functions. 
For a countable ordinal  $\alpha\geq 1$, a set $F\subseteq X^\om$ is said to be 
a ${\bf \Si}^0_\alpha$  
(respectively,  ${\bf \Pi}^0_\alpha$, ${\bf \Si}^1_1$)-{\it complete set} 
iff for any set $E\subseteq Y^\om$  (with $Y$ a finite alphabet): 
 $E\in {\bf \Si}^0_\alpha$ (respectively,  $E\in {\bf \Pi}^0_\alpha$,  $E\in {\bf \Si}^1_1$) 
iff there exists a continuous function $f: Y^\om \ra X^\om$ such that $E = f^{-1}(F)$. 

 We now   recall the definition of classes of the arithmetical hierarchy of $\om$-languages, see \cite{Staiger97}. 
Let $X$ be a finite alphabet or $X=\om$. An \ol~ $L\subseteq X^\om$  belongs to the class 
$\Si_n$ if and only if there exists a recursive relation 
$R_L\subseteq (\mathbb{N})^{n-1}\times X^\star$  such that:
\nl $~~~~~~~~~~~~~~~~~ ~~~~~~~~L = \{\sigma \in X^\om \mid \exists a_1\ldots Q_na_n  \quad (a_1,\ldots , a_{n-1}, 
\sigma[a_n+1])\in R_L \},$
\nl where $Q_i$ is one of the quantifiers $\fa$ or $\exists$ 
(not necessarily in an alternating order). An $\om$-language $L\subseteq X^\om$  belongs to the class 
$\Pi_n$ if and only if its complement $X^\om - L$  belongs to the class 
$\Si_n$.  
The  class  $\Si^1_1$ is the class of {\it effective analytic sets} which are 
 obtained by projection of arithmetical sets.
An \ol~ $L\subseteq X^\om$  belongs to the class 
$\Si_1^1$ if and only if there exists a recursive relation 
$R_L\subseteq \mathbb{N}\times \{0, 1\}^\star \times X^\star$  such that:
$$L = \{\sigma \in X^\om  \mid \exists \tau (\tau\in \{0, 1\}^\om \wedge \fa n \exists m 
 ( (n, \tau[m], \sigma[m]) \in R_L )) \}.$$
\noi 
 Then an \ol~ $L\subseteq X^\om$  is in the class $\Si_1^1$ iff it is the projection 
of an \ol~ over the alphabet $X\times \{0, 1\}$ which is in the class $\Pi_2$.  The   class $\Pi_1^1$ of  {\it effective co-analytic sets} 
 is simply the class of complements of effective analytic sets. 

Recall that the (lightface) class $\Si_1^1$ of effective analytic sets is strictly included into the (boldface) class ${\bf \Si}^1_1$ of analytic sets.

 Recall that a B\"uchi Turing machine is just a Turing machine working on infinite inputs (over a finite alphabet)  
with a B\"uchi-like acceptance condition, and 
that the class of  $\om$-languages accepted by  B\"uchi Turing machines is the class $ \Si^1_1$ of effective analytic sets  \cite{CG78b,Staiger97}.
On the other hand, one can  construct, using a classical  construction (see for instance  \cite{HopcroftMotwaniUllman2001}),  from a  B\"uchi 
Turing machine $\mathcal{T}$,  a $2$-counter  B\"uchi  automaton $\mathcal{A}$ accepting the same $\om$-language. 
Thus one can state the following proposition. 

\begin{Pro}[\cite{Staiger97,Staiger99}]\label{tm} Let $X$ be a finite alphabet. 
An \ol~ $L\subseteq X^\om$ is in the class $\Si_1^1$
iff it is accepted by a non deterministic  B\"uchi  Turing machine,  hence  iff it is in the class ${\bf BCL}(2)_\om$. 
\end{Pro}

We assume also   the reader to be familiar with the  arithmetical and  analytical hierarchies
on subsets of  $\mathbb{N}$,  these notions  may be found in the textbooks on computability theory \cite{rog} 
\cite{Odifreddi1,Odifreddi2}.

\section{Gale-Stewart games specified by 2-tape automata}

We first recall the definition of Gale-Stewart games. 

\begin{defi}[\cite{Jech}]
Let $A\subseteq X^\om$, where $X$ is a finite or countably infinite alphabet.
 The Gale-Stewart  game $G(A)$ is a game with perfect
 information between two players. Player 1 first writes a letter
$a_1\in X$, then Player 2 writes a letter $b_1\in X$,
 then Player 1 writes $a_2\in X$, and so on $\ldots$
After $\om$ steps, the two players have composed a word 
$x =a_1b_1a_2b_2\ldots$ of $X^\om$.
 Player 1 wins the play iff $x \in A$, otherwise Player 2
wins the play.

Let $A\subseteq X^\om$ and $G(A)$ be the associated Gale-Stewart  game. A strategy for Player 1 
is a function $F_1: (X^2)^\star \ra X$ and a strategy for Player 2 is a function $F_2: (X^2)^\star X  \ra X$. 
Player 1 follows the strategy $F_1$ in a play if  for each integer $n\geq 1$ ~~  $a_n = F_1(a_1b_1a_2b_2 \cdots a_{n-1}b_{n-1})$. If Player 1 
wins every play in which she has followed the strategy $F_1$, then we say that 
the strategy $F_1$ is a winning strategy (w.s.)  for Player 1.  The notion of winning strategy for Player 2 is defined in a similar manner.  

 The game $G(A)$  is said to be determined if one of the two players has a winning strategy.

 We shall denote {\bf Det}($\mathcal{C}$), where $\mathcal{C}$ is a class of $\om$-languages, 
the sentence : ``Every Gale-Stewart  game $G(A)$, where $A\subseteq X^\om$ is an $\om$-language in the class $\mathcal{C}$, is determined". 
\end{defi}

Notice that, in the whole paper, we assume that ZFC is consistent, and all results, lemmas, propositions, theorems, 
are stated in ZFC unless we explicitely give another axiomatic framework.

\hs Notice that it is known that the determinacy of effective analytic games for $X=\omega$, i.e. for a countably infinite alphabet, is equivalent to 
the determinacy of effective analytic games for a finite alphabet $X$. This follows easily from Lemma \ref{2alph} below.  
In the sequel the determinacy of effective analytic games  will be denoted by {\bf Det}($\Si_1^1$). 

\hs 
The following results were  successively proved in \cite{Fin13-JSL}. 

\begin{Pro}\label{the1}
{\bf Det}($\Si_1^1$)  $\Longleftrightarrow$ {\bf Det}({\bf r}-${\bf BCL}(8)_\om$). 
\end{Pro}

\begin{theorem}\label{the2}
{\bf Det}($\Si_1^1$)  $\Longleftrightarrow$     {\bf Det}(${\bf CFL}_\om$)            $\Longleftrightarrow$ {\bf Det}(${\bf BCL}(1)_\om$). 
\end{theorem}

\begin{theorem}\label{the3}
{\bf Det}($\Si_1^1$)  $\Longleftrightarrow$     {\bf Det}(${\bf CFL}_\om$)     
$\Longleftrightarrow$ {\bf Det}({\bf r}-${\bf BCL}(1)_\om$). 
\end{theorem}

\hs We now consider Gale-Stewart games of the form $G(A)$ where $A \subseteq X^\om$,  $X= \Si \times \Ga$ is the product of two finite alphabets, and 
$A=L(\mathcal{A})\subseteq (\Si \times \Ga)^\om$ is an infinitary rational relation
accepted by a 2-tape B\"uchi automaton $\mathcal{A}$. 

Recall that an infinite word over the alphabet   $X= \Si \times \Ga$ may be identified with a pair of infinite words $(u,v) \in \Sio \times \Ga^\om$ and so we 
often identify  $(\Si \times \Ga)^\om$ and $\Sio \times \Ga^\om$. 

We are going to prove the following result.  

\begin{theorem}\label{the4}
{\bf Det}($\Si_1^1$)  $\Longleftrightarrow$   {\bf Det}(${\bf RAT}_\om$). 
\end{theorem}

 In order to prove this result, we shall use the equivalence {\bf Det}($\Si_1^1$)  $\Longleftrightarrow$   
{\bf Det}({\bf r}-${\bf BCL}(1)_\om$) which was proved in \cite{Fin12,Fin13-JSL}.

\hs  We now first  define a coding of  an $\om$-word over a finite alphabet $\Si$, 
by an  $\om$-word over the  alphabet $\Si_1 = \Si \cup \{0, A\}$, where  $0, A$ are additional letters 
not in $\Si$. 

\hs For $x\in \Sio$  the $\om$-word $h(x)$ is defined by: 
$$h(x) = 0.Ax(1).0^2.x(2).0^3.A.x(3).0^4.x(4). \ldots 0^{2n}.x(2n).0^{2n+1}.A.x(2n+1)\ldots$$

\noi Notice that the $\om$-word $h(x)$ is obtained from the $\om$-word 
$$0.x(1).0^2.x(2).0^3.x(3).0^4.x(4)\ldots $$
by adding a letter $A$ before each letter $x(2n+1)$, where $n\geq0$ is an integer. 

\hs Let also 

$$\alpha = 0.AA.0^2.A.0^3.AA.0^4.A.0^5 \ldots AA.0^{2n}.A.0^{2n+1}.AA.0^{2n+2}\ldots$$

\noi Notice that this $\om$-word $\alpha$ is easily obtained from the $\om$-word 
$$\alpha' = 0.A.0^2.A.0^3.A.0^4.A.0^5.A \ldots A.0^n.A.0^{n+1}.A \ldots$$
by adding a letter $A$ before each segment $A.0^{2n}.A$, where $n\geq1$ is an integer.

\hs  Then it is easy to see that the mapping $h$ from $\Sio$ into $(\Si \cup \{0, A\})^\om$ is continuous and injective.

 We can now state the following Lemma.

\begin{lem}\label{R1}
Let $\Si$  be a finite alphabet and $0, A$ be two additional letters 
not in $\Si$. Let   $\alpha$ be the $\om$-word over  $\Ga=\{0, A\}$ defined as above, and 
 $L \subseteq \Sio$ be in  {\bf r}-${\bf BCL}(1)_\om$.
Then there exists  an infinitary rational relation 
$R_1 \subseteq (\Si\cup\{0,A\})^\om \times \Ga^\om$ such that:
$$\fa x\in \Si^{\om}~~~ (x\in L) \mbox{  iff } ( (h(x), \alpha) \in R_1 )$$ 
\end{lem}

\proo Let $\Si$  be a finite alphabet,  $0, A$ be two additional letters 
not in $\Si$. Let  $\alpha$ be the $\om$-word 
over  $\{0, A\}$ defined as above, and  $L = L(\mathcal{A}) \subseteq \Sio$, where 
$\mathcal{A}$=$(K,\Si, \Delta, q_0, F)$ is a real time $1$-counter B\"uchi automaton. 

\hs We  now  define the relation $R_1$.

\hs A pair 
  $y=(y_1, y_2) \in (\Si\cup\{0, A\})^\om \times \Ga^\om$ is in $R_1$ 
if and only if it is in the form

\hs $y_1 = u_1.v_1.A.x(1).u_2.v_2.x(2).u_3.v_3.A.x(3)
\ldots .u_{2n}.v_{2n}.x(2n).u_{2n+1}.v_{2n+1}.A.x(2n+1). \ldots$

\hs $y_2 = w_1.z_1.AA.w_2.z_2.A.w_3.z_3.AA \ldots 
 AAw_{2n}.z_{2n}.A.w_{2n+1}.z_{2n+1} \ldots$

\hs where $|v_1|=0$ and 
for all integers $i\geq 1$, 
$$ u_i ,v_i, w_i, z_i \in 0^\star \mbox{ and } x(i) \in \Si  \mbox{  and   }  $$ 
$$~~~~~ |u_{i+1}|=|z_i|+1 $$ 

\noi and there is a sequence $(q_i)_{i\geq 0}$  of states of $K$ 
 such that  for all integers 
$i\geq 1$:  

$$  x(i) : ( q_{i-1}, |v_i| ) \mapsto_{\mathcal{A}} 
(q_i, |w_i| )$$

\noi Moreover some state $q_f \in F$ occurs infinitely often in the sequence $(q_i)_{i\geq 0}$. 
\nl  Notice that the state $q_0$ of the sequence $(q_i)_{i\geq 0}$  is also the initial state 
of $\mathcal{A}$. 

\hs Notice that the main idea is that we try to simulate, using a $2$-tape automaton,  the reading of the infinite word $x(1).x(2).x(3) \ldots$ by  the 
real time $1$-counter B\"uchi automaton $\mathcal{A}$. The initial value of the counter is $|v_1|$ and the value  of the counter 
after the reading of the letter $x(1)$ by  $\mathcal{A}$  is $|w_1|$ which is  on the second tape. Now the $2$-tape automaton accepting 
$R_1$ would need to read again the value $|w_1|$ 
in order to compare it to the value of the counter after the reading of $x(2)$ by the $1$-counter  automaton $\mathcal{A}$. 
This is not directly possible so the simulation does not work on every pair of $R_1$. However,  using the very special shape of  pairs in 
$h(\Sio) \times \{\alpha\}$, the simulation will be possible on a pair $(h(x), \alpha)$. Then for such a pair  $ (h(x), \alpha) \in R_1$ written 
in the above form $(y_1, y_2)$,  we have $|v_2|=|w_1|$ and then  the simulation can continue from the value $|v_2|$ of the counter, and so on.

\hs We now give the details of the proof. 
\nl Let  $x\in \Si^{\om}$ be  such that  $(h(x), \alpha) \in R_1$. We are going to prove that $x\in L$. 

\hs  By hypothesis  $ (h(x), \alpha) \in R_1$ thus there are finite words  $u_i ,v_i, w_i, z_i \in 0^\star$ such that 
 $|v_1|=0$ and for all integers $i\geq 1$, $|u_{i+1}|=|z_i|+1 $, and  

\hs $y_1 = u_1.v_1.A.x(1).u_2.v_2.x(2).u_3.v_3.A.x(3)
\ldots .u_{2n}.v_{2n}.x(2n).u_{2n+1}.v_{2n+1}.A.x(2n+1). \ldots$

\hs $y_2 = w_1.z_1.AA.w_2.z_2.A.w_3.z_3.AA \ldots 
 AAw_{2n}.z_{2n}.A.w_{2n+1}.z_{2n+1} \ldots$

\hs Moreover  there is a sequence $(q_i)_{i\geq 0}$  of states of $K$ 
 such that  for all integers 
$i\geq 1$:  

$$  x(i) : ( q_{i-1}, |v_i| ) \mapsto_{\mathcal{A}} 
(q_i, |w_i| )$$

\noi and  some state $q_f \in F$ occurs infinitely often in the sequence $(q_i)_{i\geq 0}$. 

\hs 
On the other side we have: 
\nl  $h(x) =  0.Ax(1).0^2.x(2).0^3.A.x(3).0^4.x(4). \ldots 0^{2n}.x(2n).0^{2n+1}.A.x(2n+1)\ldots$
\nl $\alpha =0.AA.0^2.A.0^3.AA.0^4.A.0^5 \ldots AA.0^{2n}.A.0^{2n+1}.AA.0^{2n+2}\ldots$

\hs So we have $|u_1.v_1|=1$ and $|v_1|=0$ and $x(1): ( q_{0}, |v_1| ) \mapsto_{\mathcal{A}} 
(q_1, |w_1| )$. But $|w_1.z_1|=1$,  $|u_2.v_2|=2$, and $|u_2|=|z_1|+1$ thus $|v_2|=|w_1|$. 

\hs We are going to prove  in a similar way  that for all integers $i\geq 1$ it holds that $|v_{i+1}|=|w_i|$. 
\nl We know that $|w_i.z_i|=i$, $|u_{i+1}.v_{i+1}|=i+1$, and $|u_{i+1}|=|z_i|+1$ thus $|w_i|=|v_{i+1}|$. 

\hs Then for all $i\geq 1$,  $x(i) : ( q_{i-1}, |v_i| ) \mapsto_{\mathcal{A}} (q_i, |v_{i+1}| )$. 
\nl So if we set $c_i=|v_i|$, $(q_{i-1}, c_{i})_{i\geq 1}$ is an accepting run of $\mathcal{A}$ on $x$  and this implies that  
$x\in L$. 
\nl Conversely it is easy to prove that if $x\in L$ then $(h(x), \alpha)$ may be written in the form of $(y_1, y_2)\in R_1$. 

\hs It remains  to prove that the above defined relation $R_1$ is an infinitary rational 
relation.  It is easy to find a $2$-tape  B\"uchi automaton $\mathcal{A}$ accepting  
the  relation $R_1$.            
\ep

\begin{lem}\label{complement} The set 
$$R_2 = (\Si\cup\{0, A\})^\om\times \Ga^\om - ( h(\Si^{\om}) \times \{\alpha\} )$$
\noi is an infinitary rational relation.  
\end{lem}

\proo By definition of the mapping $h$, we know that a pair of $\om$-words $(\sigma_1, \sigma_2)$
 is in $h(\Si^{\om}) \times \{\alpha\}$ iff 
 it is of the  form: 

\hs  $\sigma_1 = h(x) =  0.Ax(1).0^2.x(2).0^3.A.x(3).0^4.x(4). \ldots 0^{2n}.x(2n).0^{2n+1}.A.x(2n+1)\ldots$

\hs  $\sigma_2 = \alpha = 0.AA.0^2.A.0^3.AA.0^4.A.0^5 \ldots AA.0^{2n}.A.0^{2n+1}.AA.0^{2n+2}\ldots$

\hs where for all integers $i\geq 1$,  $x(i)\in \Si$. 

\hs So it is easy to see that 
$(\Si\cup\{0, A\})^\om\times \Ga^\om - ( h(\Si^{\om}) \times \{\alpha\} )$
is the union of the sets $\mathcal{C}_j$ where:

\begin{itemize} 

\ite $\mathcal{C}_1$ is formed by pairs  $(\sigma_1, \sigma_2)$ where 
\nl $\sigma_1$ has not any initial segment in $0.A.\Si.0^2.\Si.0^3A.\Si$, 
or 
\nl $\sigma_2$ has not any initial segment in $0.AA.0^2.A.0^3AA$.

\ite $\mathcal{C}_2$ is formed by pairs  $(\sigma_1, \sigma_2)$ where 
\nl $\sigma_2 \notin  (0^+AA0^+A)^\om$, or 
\nl $\sigma_1 \notin (0^+.A.\Si.0^+.\Si)^\om$.  

\ite $\mathcal{C}_3$ is formed by pairs  $(\sigma_1, \sigma_2)$ where 
\nl $\sigma_1 = w_1.u.A.z_1 $
\nl $\sigma_2 = w_2.v.A.z_2 $
\hs where $n$ is an integer $\geq 1$,  $w_1\in (0^+.A.\Si.0^+.\Si)^n$, $w_2\in (0^+AA0^+A)^n$, 
\nl $u, v \in 0^+$, $z_1\in (\Si\cup\{0, A\})^\om$, $z_2 \in \Ga^\om$, and 
$$ |u| \neq |v|$$

\ite $\mathcal{C}_4$ is formed by pairs  $(\sigma_1, \sigma_2)$ where 
\nl $\sigma_1 = w_1.u.z_1 $
\nl $\sigma_2 = w_2.v.A.z_2 $

\hs where $n$ is an integer $\geq 1$,  
\nl $w_1\in (0^+.A.\Si.0^+.\Si)^n.0^+.A.\Si.$, 
\nl $w_2\in (0^+AA0^+A)^n.0^+AA$, 
\nl $u, v \in 0^+$, $z_1\in \Si.(\Si\cup\{0, A\})^\om$, $z_2 \in \Ga^\om$, and 
$$ |u| \neq |v|$$

\ite $\mathcal{C}_5$ is formed by pairs  $(\sigma_1, \sigma_2)$ where 
\nl $\sigma_1 = w_1.u.A.b.w.c.A.z_1 $
\nl $\sigma_2 = w_2.v.A.z_2 $
\hs where $n$ is an integer $\geq 1$,  

\hs where $n$ is an integer $\geq 1$,  
$w_1\in (0^+.A.\Si.0^+.\Si)^n$, $w_2\in (0^+AA0^+A)^n$, 
\nl $u, v, w \in 0^+$, $b, c \in \Si$,  $z_1\in (\Si\cup\{0, A\})^\om$, $z_2 \in \Ga^\om$, and 
$$ |w| \neq |v| +1 $$

\ite $\mathcal{C}_6$ is formed by pairs  $(\sigma_1, \sigma_2)$ where 
\nl $\sigma_1 = w_1.u.A.b.w.c.w''.A.z_1 $
\nl $\sigma_2 = w_2.v.AA.w'.A z_2 $
\hs where $n$ is an integer $\geq 1$,  

\hs where $n$ is an integer $\geq 1$,  $w_1\in (0^+.A.\Si.0^+.\Si)^n$, $w_2\in (0^+AA0^+A)^n$, 
\nl $u, v, w, w', w'' \in 0^+$, $b, c \in \Si$,  $z_1\in (\Si\cup\{0, A\})^\om$, $z_2 \in \Ga^\om$, and 
$$  |w''| \neq |w'| +1  $$

\end{itemize}

\noi It is easy to see that for each integer  $j \in [1, 6]$, the set  $\mathcal{C}_j \subseteq (\Si\cup\{0, A\})^\om\times \Ga^\om$ is  an infinitary 
rational relation.  The class ${\bf RAT}_\om$ is closed under finite union thus

$$R_2 = (\Si\cup\{0, A\})^\om\times \Ga^\om - ( h(\Si^{\om}) \times \{\alpha\} )
 = \bigcup_{1\leq j\leq 6} \mathcal{C}_j$$

\noi is an infinitary rational relation. \ep  

\hs \noi {\bf End of Proof of Theorem \ref{the4}.} 
\nl The implication {\bf Det}($\Si_1^1$)  $\Longrightarrow$   {\bf Det}(${\bf RAT}_\om$)  follows directly from the inclusion 
${\bf RAT}_\om \subseteq \Si_1^1$. 

\hs To prove the reverse  implication  {\bf Det}(${\bf RAT}_\om$)  $\Longrightarrow$   {\bf Det}($\Si_1^1$), 
we assume that {\bf Det}(${\bf RAT}_\om$)   holds and   we  show that   
every Gale-Stewart game $G(L)$, where $L\subseteq \Si^\om$ is an $\om$-language in the class {\bf r}-${\bf BCL}(1)_\om$ is determined.  
Then Theorem  \ref{the3} will imply that {\bf Det}($\Si_1^1$)  also holds. 

 Let then $L= L(\mathcal{A})\subseteq \Si^\om$ be an $\om$-language in the class {\bf r}-${\bf BCL}(1)_\om$ which is accepted by a real-time 
$1$-counter B\"uchi automaton $\mathcal{A}$=$(K,\Si, \Delta, q_0, F)$.   

We shall consider a Gale-Stewart game $G(\mathcal{L})$ where $\mathcal{L}\subseteq (\Si\cup\{0, A\})^\om\times \Ga^\om$, the letters  
$0, A$ are  not in $\Si$ and $\Ga=\{0, A\}$,   and we are going to 
define a suitable winning set $\mathcal{L}$ accepted by a $2$-tape B\"uchi automaton. 

Notice first that in such a game, the players alternatively write letters $(a_i, b_i)$, $i\geq 1$, 
from the product alphabet $X=(\Si\cup\{0, A\}) \times \Ga$. After $\om$ steps 
they have produced an $\om$-word $y \in X^\om$ where $y$ may be identified with a pair $(y_1, y_2)\in (\Si\cup\{0, A\})^\om\times \Ga^\om$. 

Consider now the coding defined above with the function $h: \Sio \ra (\Si\cup\{0, A\})^\om$, and the $\om$-word $\alpha\in  \Ga^\om$. 
This coding is inspired from a previous one we have used to study the topological complexity of infinitary rational relations \cite{Fin06b,Fink-Wd}. 
We have here modified this previous coding to get some useful properties for the game we are going to define. 

Assume that two players 
alternatively write letters from the alphabet $X=(\Si\cup\{0, A\}) \times \Ga$
and that they finally produce an $\om$-word in the form $y=(h(x), \alpha)$ for some $x\in \Sio$.  
We now have  the two following properties which will be useful in the sequel. 

(1) The letters $x(2n+1)$, for $n\geq 0$,  have been  written by Player 1, and the letters $x(2n)$, for $n\geq 1$, have been  written by Player 2. 

(2) After a sequence of consecutive letters $0$, either on the first component $h(x)$ or on the second component $\alpha$, 
 the first letter which is not a $0$ has always been  written by Player 2. 

This  is due in particular to the following fact:  the  sequences of letters $0$ 
 on the first component $h(x)$ or on the second component $\alpha$ are alternatively of odd and even lengths. 

On the other hand we can remark that all $\om$-words in the form $h(x)$ belong to the $\om$-language $H \subseteq (\Si\cup\{0, A\})^\om$ 
defined by: 

$$H = [ (0^2)^\star.0.A.\Si.(0^2)^+.\Si ] ^\om$$

In a similar way the $\om$-word $\alpha$ belongs to the $\om$-language $H' \subseteq \Ga^\om$ defined by: 

$$H' = [ (0^2)^\star.0.AA.(0^2)^+.A ] ^\om$$

\hs An important fact is the following property of $H \times H'$  which extends the same property of  the set $h(\Sio)\times \{\alpha\}$. 
Assume that two players 
alternatively write letters from the alphabet $X=(\Si\cup\{0, A\}) \times \Ga$
and that they finally produce an $\om$-word $y=(y_1, y_2)$ in $H\times H'$ in the following form: 

\hs  $y_1 =   0^{n_1}.Ax(1).0^{n_2}.x(2).0^{n_3}.A.x(3).0^{n_4}.x(4). \ldots   0^{n_{2k}}.x(2k).0^{n_{2k+1}}.A.x(2k+1)\ldots$

\hs  $y_2 = \alpha = 0^{n'_1}.AA.0^{n'_2}.A.0^{n'_3}.AA.0^{n'_4}.A.0^{n'_5} \ldots AA.0^{n'_{2k}}.A.0^{n'_{2k+1}}.AA.0^{n'_{2k+2}}\ldots$

\hs where for all integers $i\geq 1$, $n_i \geq 1$ (respectively, $n'_i$)   
is an odd integer iff $i$ is an odd integer and $n_i$ (respectively, $n'_i$) is an even integer iff $i$ is an  even integer. 

\hs   Then we have the two following facts: 

(1)  The letters $x(2n+1)$, for $n\geq 0$,  have been  written by Player 1, and the letters $x(2n)$, for $n\geq 1$, have been  written by Player 2. 

(2)  After a sequence of consecutive letters $0$  (either on the first component $y_1$ or on the second component $y_2$),   
 the first letter which is not a $0$  has always been  written by Player 2.

\hs   Let now 
$$V={\rm Pref}(H) \cap (\Si\cup\{0, A\})^\star .0$$

\noi  So a finite word over   the alphabet $\Si\cup\{0, A\}$ is in $V$ iff it is a 
prefix  of some word in  $H$ and its  last letter is a $0$. It is easy to see that the topological closure of $H$ is 
${\rm Cl}(H)=H ~   \cup~    V.0^\om.$

\hs In a similar manner let 
$$V'={\rm Pref}(H') \cap (\Ga)^\star .0$$
\noi  So a finite word over   the alphabet $\Ga$ is in $V'$ iff it is a 
prefix  of some word in  $H'$ and its  last letter is a $0$. It is easy to see that the topological closure of $H'$ is 
${\rm Cl}(H')=H' ~   \cup~    V'.0^\om. $

\hs Notice that an $\om$-word $x$ in ${\rm Cl}(H)$ is not in $h(\Si^\om)$ iff a sequence of consecutive letters $0$ in $x$ has not the good length. 
And  an $\om$-word $y$ in ${\rm Cl}(H')$ is not equal to $\alpha$ iff a sequence of consecutive letters $0$ in $y$ has not the good length. 

\hs Thus if 
two players alternatively write letters from the alphabet $X=(\Si\cup\{0, A\}) \times \Ga$
and that they finally produce an $\om$-word in the form $y = (y_1, y_2) \in {\rm Cl}(H) \times {\rm Cl}(H')- h(\Si^\om) \times \{\alpha\}$ then 
it is Player 2 who ``has gone out" of the {\it closed} set $h(\Si^\om) \times \{\alpha\}$ at some step of the play. 
This means that there is an integer $n\geq 1$  such that $y[2n-1]\in {\rm Pref}(h(\Si^\om) \times \{\alpha\})$ and 
$y[2n] \notin {\rm Pref}(h(\Si^\om) \times \{\alpha\})$. In a similar way we shall say that, during  an infinite play, Player 1 ``goes out" 
 of the {\it closed} set $h(\Si^\om) \times \{\alpha\}$ if the final  play $y$ composed by the two players has a prefix $y[2n]\in {\rm Pref}(h(\Si^\om) \times \{\alpha\})$ such that 
$y[2n+1]\notin {\rm Pref}(h(\Si^\om) \times \{\alpha\})$. 
This will be important in the sequel.

\hs From Lemmas \ref{R1} and \ref{complement} we know that we can effectively construct a $2$-tape B\"uchi automaton $\mathcal{B}$
such that 
$$L(\mathcal{B}) = [ h(L(\mathcal{A}))  \times \{\alpha\} ]  \cup  [( h(\Si^{\om}) \times \{\alpha\})^- ]$$

\hs  On the other hand it is very easy to see that the $\om$-language $H$ (respectively, $H'$) is regular 
and to construct a B\"uchi automaton $\mathcal{H}$ 
 (respectively, $\mathcal{H}'$)  accepting it. Therefore one can also construct a $2$-tape B\"uchi automaton $\mathcal{B}'$ 
such that 
$$L(\mathcal{B}') = [h(L(\mathcal{A}))  \times \{\alpha\} ] \cup  [( h(\Si^{\om}) \times \{\alpha\})^- \cap H \times H'] $$

\noi Notice also that ${\rm Pref}(H)$ (respectively,  ${\rm Pref}(H')$)   
 is a regular finitary language since $H$ (respectively, $H'$)  is a regular $\om$-language.  Thus the $\om$-languages $V.0^\om$ and $V'.0^\om$ 
are also regular.  Moreover the closure of a regular $\om$-language is a regular $\om$-language  thus ${\rm Cl}(H)$ and ${\rm Cl}(H')$ are also regular, and 
we can construct, from the B\"uchi automata $\mathcal{H}$ and  $\mathcal{H}'$, some other B\"uchi automata $\mathcal{H}_c$ and  $\mathcal{H}_c'$
acccepting the regular $\om$-languages ${\rm Cl}(H)$ and ${\rm Cl}(H')$, 
\cite{PerrinPin}. 
Thus  one can construct a $2$-tape B\"uchi automaton $\mathcal{C}$ such that: 

$$L(\mathcal{C})= [  V.0^\om \times {\rm Cl}(H') ] ~ \cup ~ [ {\rm Cl}(H) \times V'.0^\om ]$$

\hs We denote also $U$ the set of finite words $u$ over $X = (\Si\cup\{0, A\})\times \Ga$ such that $|u|=2n$ for some integer $n\geq 1$ 
and $u[2n-1] \in {\rm Pref}(H) \times {\rm Pref}(H')$ and 
$u=u[2n] \notin {\rm Pref}(H)\times {\rm Pref}(H')$. Since the regular languages ${\rm Pref}(H)$ and ${\rm Pref}(H')$ are 
accepted by finite automata, one can construct 
a $2$-tape B\"uchi automaton $\mathcal{C'}$ such that: 
$$L(\mathcal{C}')=  U.[ (\Si\cup\{0, A\})^\om \times \Gao ] $$

Now we set: 
$$\mathcal{L}~ ~   =  ~ ~  L(\mathcal{B}')~ ~  \cup ~ ~   L(\mathcal{C}) ~ ~   \cup~ ~ L(\mathcal{C}')  $$

i.e. 

$$\mathcal{L}~ ~   =  ~ ~[  h(L(\mathcal{A}))  \times \{\alpha\} ] ~ ~  \cup  ~ ~  [( h(\Si^{\om}) \times \{\alpha\})^- \cap H \times H']  ~ ~  
\cup ~ ~   L(\mathcal{C}) ~ ~   \cup~ ~ L(\mathcal{C}')  $$

The class of infinitary rational relations is effectively closed under finite union, thus we  can construct a $2$-tape B\"uchi automaton 
$\mathcal{D}$ such that  $\mathcal{L}=L(\mathcal{D})$.

\hs By hypothesis  we assume that {\bf Det}(${\bf RAT}_\om$) holds and thus the game $G(\mathcal{L})$ is determined. We are going to show that 
this implies that the game $G( L(\mathcal{A}) )$ itself is determined. 

\hs Assume firstly that Player 1 has a winning strategy $F_1$  in the game $G(\mathcal{L})$.  

If during an infinite play, the two players compose an infinite word $z \in X^\om$, and  Player 2 ``does not go out of the set
 $h(\Si^{\om})\times \{\alpha\}$" 
then we claim that also Player 1, following her strategy $F_1$, ``does not go out  of the set $h(\Si^{\om})\times \{\alpha\}$". 
Indeed if Player 1 goes out  of this set then due to the above remark this would imply that Player 1 also goes out of the set 
${\rm Cl}(H) \times {\rm Cl}(H')$: there is an integer $n\geq 0$ such that $z[2n] \in {\rm Pref}(H \times H')$ but  $z[2n+1] \notin {\rm Pref}(H \times H')$. 
So  $z \notin h(L(\mathcal{A}))  \times \{\alpha\} ~ ~  \cup  ~ ~  [( h(\Si^{\om}) \times \{\alpha\})^- \cap H \times H']  ~ ~  
\cup ~ ~   L(\mathcal{C}) $.  
Moreover it follows from  the definition of $U$  that $z \notin  L(\mathcal{C}')=  U.[ (\Si\cup\{0, A\})^\om \times \Gao ]$. 
Thus If Player 1  goes out  of the set $h(\Si^{\om})\times \{\alpha\}$ then she looses the game. 

Consider now  an infinite play in which   Player 2 ``does not go out of the set $h(\Si^{\om})\times \{\alpha\}$".  
Then Player 1, following her strategy $F_1$, ``does not go out  of the set $h(\Si^{\om})\times \{\alpha\}$".  
Thus the two players write an infinite word $z=(h(x),\alpha)$ for some 
infinite word $x\in \Si^{\om}$. But the  letters $x(2n+1)$, for $n\geq 0$,  have been  written by Player 1, and the letters $x(2n)$, for $n\geq 1$, 
have been  written by Player 2.  Player 1 wins the play iff $x\in L(\mathcal{A})$  and Player 1 wins always the play when she uses her strategy $F_1$. 
This implies that Player 1 has also a w.s. in the game  $G( L(\mathcal{A}) )$. 

\hs Assume now that Player 2 has a winning strategy $F_2$  in the game $G(\mathcal{L})$. 
 
If during an infinite play, the two players compose an infinite word $z$, and  Player 1 ``does not go out of the set $h(\Si^{\om})\times \{\alpha\}$" 
then we claim that also Player 2, following his strategy $F_2$, ``does not go out  of the set $h(\Si^{\om})\times \{\alpha\}$". Indeed if 
Player 2 goes out   of the set $h(\Si^{\om})\times \{\alpha\}$ and the final play $z$ remains in 
${\rm Cl}(H\times H')={\rm Cl}(H) \times {\rm Cl}(H')$ then 
$z\in  [( h(\Si^{\om}) \times \{\alpha\})^- \cap H \times H']  ~ ~  
\cup ~ ~   L(\mathcal{C})  \subseteq \mathcal{L}$ and 
Player 2 looses.  If Player 1 does not go out of the set ${\rm Cl}(H \times H')$ and 
at some step of the play, Player 2 goes out of ${\rm Cl}(H) \times {\rm Cl}(H')$, i.e. there is an integer $n\geq 1$ such that $z[2n-1] 
\in {\rm Pref}(H)\times {\rm Pref}(H')$
and $z[2n] \notin {\rm Pref}(H)\times {\rm Pref}(H')$,  then $z \in U.[ (\Si\cup\{0, A\})^\om \times \Gao ]  \subseteq \mathcal{L}$ and Player 2 looses. 

Assume now that Player 1 ``does not go out of the set $h(\Si^{\om})\times \{\alpha\}$". Then  Player 2 follows his w. s.  $F_2$, 
and then ``never goes out  of the set $h(\Si^{\om})\times \{\alpha\}$". 
 Thus the two players write an infinite word $z=(h(x), \alpha)$ for some 
infinite word $x\in \Si^{\om}$. But the  letters $x(2n+1)$, for $n\geq 0$,  have been  written by Player 1, and the letters $x(2n)$, for $n\geq 1$, 
have been  written by Player 2.  Player 2  wins the play iff $x\notin L(\mathcal{A})$  and Player 2 wins always the play when he uses his strategy $F_2$. 
This implies that Player 2 has also a w.s. in the game  $G( L(\mathcal{A}) )$. 
\ep

\hs Recall the following effective  result cited in \cite[remark 3.5]{Fin13-JSL} which follows from the proofs of Proposition 
 \ref{the1}  and Theorems \ref{the2} and  \ref{the3}. 

\begin{Pro}\label{pro-r1c}
Let $L\subseteq X^\om$ be an $\om$-language in the class $\Si_1^1$, or equivalently in the class 
${\bf BCL}(2)_\om$, which is accepted by a B\"uchi $2$-counter automaton $\mathcal{A}$. Then one can effectively construct from $\mathcal{A}$
a real time B\"uchi $1$-counter automaton $\mathcal{B}$ such that the game $G(L)$  is determined if and only if the game  $G(L(\mathcal{B}))$ is 
determined. Moreover Player 1 (respectively, Player 2)  has a w.s. in the game $G(L)$  iff  Player 1 (respectively, Player 2) 
 has a w.s. in the game 
$G(L(\mathcal{B}))$. 
\end{Pro}

\noi We can easily see, from the proofs of  Proposition 
 \ref{the1}  and Theorems \ref{the2} and  \ref{the3} in  \cite{Fin13-JSL},  that we have also the following additional property 
which strengthens the above one.

\begin{Pro}\label{pro-recws} 
With the same notations as in the above Proposition, if $\sigma$ is a winning strategy for Player 1 (respectively, Player 2)  in the game $G(L)$  
then one can construct a w.s. $\sigma'$ for Player 1 (respectively, Player 2)  in the game $G(L(\mathcal{B}))$ such that $\sigma'$ is recursive in $\sigma$.  
And conversely, if     $\sigma$ is a winning strategy for Player 1 (respectively, Player 2)  in the game   $G(L(\mathcal{B}))$ then one                                        
can construct a w.s. $\sigma'$ for Player 1 (respectively, Player 2)  in the game $G(L)$  such that $\sigma'$ is recursive in $\sigma$.  
\end{Pro}

Moreover we can easily see, from the proof of the above Theorem \ref{the4}, that we have also the following property. 

\begin{Pro}\label{pro-2t} 
Let $\mathcal{A}$ be a real time B\"uchi $1$-counter automaton. Then one can effectively construct from $\mathcal{A}$ a $2$-tape B\"uchi automaton
$\mathcal{B}$ such that the game $G(L(\mathcal{A}))$  is determined if and only if the game  $G(L(\mathcal{B}))$ is 
determined. Moreover Player 1 (respectively, Player 2)  has a w.s. in the game $G(L(\mathcal{A}))$  iff  Player 1 (respectively, Player 2) 
 has a w.s. in the game  $G(L(\mathcal{B}))$ and if $\sigma$ is a winning strategy for Player 1 (respectively, Player 2)  in the game 
$G(L(\mathcal{A}))$  then one can construct a w.s. $\sigma'$ for 
Player 1 (respectively, Player 2)  in the game $G(L(\mathcal{B}))$ such that $\sigma'$ is recursive in $\sigma$.  And similarly  
if $\sigma$ is a winning strategy for Player 1 (respectively, Player 2)  in the game 
$G(L(\mathcal{B}))$  then one can construct a w.s. $\sigma'$ for 
Player 1 (respectively, Player 2)  in the game $G(L(\mathcal{A}))$ such that $\sigma'$ is recursive in $\sigma$.
\end{Pro}

\noi   Recall that, assuming that ZFC is consistent, there are some models of ZFC in which {\bf Det}($\Si_1^1$) does not hold. Therefore there 
are some models of  ZFC in which some Gale-Stewart games $G(L(\mathcal{A}))$, where $\mathcal{A}$ is a one-counter B\"uchi automaton or a $2$-tape  B\"uchi automaton, 
are not determined. 

\hs Some very  natural questions now  arise. 

\hs \noi {\bf Question 1.} If we live in a model of ZFC in which {\bf Det}($\Si_1^1$)  holds, then all Gale-Stewart games $G(L(\mathcal{A}))$, 
where $\mathcal{A}$ is a one-counter B\"uchi automaton or a $2$-tape  B\"uchi automaton, are determined. Is it then possible to construct the winning strategies 
in an effective way ? 

 \hs  \noi {\bf Question 2.}  We know from Martin's Theorem that in any model of ZFC the Gale-Stewart Borel games are determined. Is it possible to 
construct effectively the winning strategies in games  $G(L(\mathcal{A}))$, when $L(\mathcal{A})$ is a Borel set, or even a Borel set of low Borel rank ?

\hs We are going to give some answers to these questions. We now firstly  recall some basic notions of set theory 
which will be useful in the sequel, and which are exposed in any  textbook on set theory, like \cite{Jech}. 

 The usual axiomatic system  ZFC is 
Zermelo-Fraenkel system  ZF   plus the axiom of choice AC. 
 The axioms of  ZFC express some  natural facts that we consider to hold in the universe of sets. For instance a natural fact is that 
two sets $x$ and $y$ are equal iff they have the same elements. 
This is expressed by the {\it Axiom of Extensionality}: 
$$\fa x \fa y ~ [ ~ x=y \leftrightarrow \fa z ( z\in x \leftrightarrow z\in y ) ~].$$
\noi   Another natural axiom is the {\it Pairing Axiom}   which states that for all sets $x$ and $y$ there exists a  set $z=\{x, y\}$ 
whose elements are $x$ and $y$: 
 $$\fa x \fa y ~ [ ~\exists z ( \fa w  ( w\in z \leftrightarrow (w=x \vee w=y) ) ) ]$$
\noi Similarly the {\it Powerset Axiom} states the existence of the set of subsets of a set $x$. 
Notice that these axioms are first-order sentences in the usual logical language of set theory whose only non logical  
symbol is the membership binary relation symbol $\in$. 
We refer the reader to any textbook on set theory  for an exposition of the other axioms of  ZFC.  

A model ({\bf V}, $\in)$ of  an arbitrary set of axioms $\mathbb{A}$  is a collection  {\bf V} of sets,  equipped with 
the membership relation $\in$, where ``$x \in y$" means that the set $x$ is an element of the set $y$, which satisfies the axioms of   $\mathbb{A}$. 
We  often say `` the model {\bf V}" instead of "the model ({\bf V}, $\in)$".

We say that two sets $A$ and $B$ have same cardinality iff there is a bijection from $A$ onto $B$ and we denote this  by $A \approx B$. 
The relation $\approx$ is an equivalence relation. 
Using the axiom of choice AC, one can prove that any set $A$ can be well-ordered and thus   there is an ordinal $\gamma$ such that $A \approx \gamma$. 
In set theory the cardinal of the set $A$ is then formally defined as the smallest such ordinal $\gamma$.

 The infinite cardinals are usually denoted by
$\aleph_0, \aleph_1, \aleph_2, \ldots , \aleph_\alpha, \ldots$
The cardinal $\aleph_\alpha$ is also denoted by $\om_\alpha$,
 when it is considered as an ordinal.
The first uncountable ordinal is $\om_1$, 
 and formally  $\aleph_1=\om_1$. The ordinal  $\om_2$ is the first ordinal of cardinality 
greater than $\aleph_1$, and so on.

Let ${\bf ON}$ be the class of all ordinals. Recall that an ordinal $\alpha$ is said to be a successor ordinal iff there exists an ordinal $\beta$ such that 
$\alpha=\beta + 1$; otherwise the ordinal $\alpha$ is said to be a limit ordinal and in this case 
$\alpha ={\rm sup} \{ \beta \in {\bf ON}\mid \beta < \alpha \}$.

\hs  The  class ${\bf L}$ of  {\it constructible sets} in a model {\bf V} of  ZF is defined by ~~~~
${\bf L} = \bigcup_{\alpha \in {\bf ON}} {\bf L}(\alpha) $, 
\noi where the sets ${\bf L}(\alpha) $ are constructed  by induction as follows: 

(1). ${\bf L}(0) =\emptyset$

(2). ${\bf L}(\alpha) = \bigcup_{\beta <  \alpha} {\bf L}(\beta) $, for $\alpha$ a limit ordinal, and 

(3).  ${\bf L}(\alpha + 1) $ is the set of subsets of ${\bf L}(\alpha) $ which are definable from a finite number of elements of ${\bf L}(\alpha) $
by a first-order formula relativized to ${\bf L}(\alpha) $.

  If  {\bf V} is  a model of  ZF and ${\bf L}$ is  the class of  {\it constructible sets} of   {\bf V}, then the class  ${\bf L}$    is a model of  
ZFC.
Notice that the axiom ( V=L), which  means ``every set is constructible",   is consistent with ZFC  because   ${\bf L}$ is a model of 
ZFC + V=L. 

 Consider now a model {\bf V} of  ZFC and the class of its constructible sets ${\bf L} \subseteq {\bf V}$ which is another 
model of  ZFC.  It is known that 
the ordinals of {\bf L} are also the ordinals of  {\bf V}, but the cardinals  in  {\bf V}  may be different from the cardinals in {\bf L}. 

  In particular,  the first uncountable cardinal in {\bf L}  is denoted 
 $\aleph_1^{\bf L}$, and it is in fact an ordinal of {\bf V} which is denoted $\om_1^{\bf L}$. 
  It is well-known that in general this ordinal satisfies the inequality 
$\om_1^{\bf L} \leq \om_1$.  In a model {\bf V} of  the axiomatic system  ZFC + V=L the equality $\om_1^{\bf L} = \om_1$ holds, but in 
some other models of  ZFC the inequality may be strict and then $\om_1^{\bf L} < \om_1$: 
notice that in this case $\om_1^{\bf L} < \om_1$ holds because there is actually a bijection from $\om$ onto $\om_1^{\bf L}$ in {\bf V} 
(so  $\om_1^{\bf L}$ is countable in {\bf V})  but no such bijection exists in the inner model {\bf L} (so $\om_1^{\bf L}$ is uncountable in {\bf L}). 
The construction of such a model is presented in \cite[page 202]{Jech}: one can start 
 from a model 
{\bf V} of  ZFC + V=L and construct by  forcing  a generic extension {\bf V[G]} in which 
$\om_1^{{\bf V}}$ is collapsed to 
 $\om$; in this extension the inequality $\om_1^{\bf L} < \om_1$ holds. 

\hs We can now state the following result, which gives an answer to Question 1.

\begin{theorem}\label{mod-zfc}
There  exists a  real-time $1$-counter 
B\"uchi automaton  $\mathcal{A}$ and a $2$-tape  B\"uchi automaton  $\mathcal{B}$  such that: 
\begin{enumerate}
\ite   There is a  model $V_1$ of  \ {\rm ZFC}  in which Player 1 has a winning strategy $\sigma$ in the  
game $G(L(\mathcal{A}))$ (respectively, $G(L(\mathcal{B}))$). 
But $\sigma$ cannot be recursive and not even in the class $(\Sigma_2^1 \cup \Pi_2^1)$. 

\ite  There is a  model  $V_2$ of  \ {\rm ZFC} in which the game $G(L(\mathcal{A}))$ (respectively, $G(L(\mathcal{B}))$)  is not determined. 
\end{enumerate}

\noi 
Moreover these are the only two possibilities: there are no models of  ZFC in which Player 2 has a winning strategy in the 
game $G(L(\mathcal{A}))$ (respectively, $G(L(\mathcal{B}))$). 
\end{theorem}

\noi To prove this result, we shall use some set theory, a result of Stern in \cite{Stern82} on coanalytic games, and the Shoenfield Absolutenesss Theorem. 

\hs We first recall Stern's result. 

\begin{theorem}[Stern \cite{Stern82}]\label{theo-stern}
For every  recursive ordinal $\xi$ there exists an effective coanalytic set $L_\xi \subseteq \om^\om$ such that 
the Gale-Stewart game $G(L_\xi)$ is determined if and only if  the ordinal  $\aleph_\xi^{\bf  L}$ is countable. Moreover if 
the game  $G(L_\xi)$ is determined then  Player 2  has a winning strategy (and thus Player 1 cannot have a w.s. in this game). 
\end{theorem}

\noi We also state the following lemmas. 

\begin{lem}\label{co-an}
Let $L \subseteq \om^\om$ be an effective coanalytic subset of the Baire space. Then there is an effective analytic subset 
$L' \subseteq \om^\om$ such that Player 1 (respectively, Player 2)  has a  w.s. in the game $G(L)$ iff Player 2 
(respectively, Player 1)  has a  w.s. in the game $G(L')$. In particular,   the game $G(L)$ is determined iff the game $G(L')$
is determined. 
\end{lem}

\proo   As noticed for instance in \cite{McAloon79},  we can associate to every effective coanalytic set  $L \subseteq \om^\om$ 
the effective analytic set $L' \subseteq \om^\om$  which is the complement of the set $L+1$ defined by: 
$$ L+1 = \{ x \in \om^\om \mid   \exists y   ~ [ y \in L  \mbox{ and }  \fa n\geq 1 ~ x(n+1)=y(n) ] \}. $$

\noi It is then easy to see that Player 1 (respectively, Player 2)  has a  w.s. in the game $G(L)$ iff Player 2 
(respectively, Player 1)  has a  w.s. in the game $G(L')$. 
\ep

\begin{lem}\label{2alph}
Let $L \subseteq \om^\om$ be an effective analytic subset of the Baire space. Then there exists an effective analytic set $L'  \subseteq \{0, 1\}^\om$
such that Player 1 (respectively, Player 2)  has a  w.s. in the game $G(L)$ iff Player 1
(respectively, Player 2)  has a  w.s. in the game $G(L')$. In particular,   the game $G(L)$ is determined iff the game $G(L')$
is determined. 
If  $L$ is an (effective)  $\Sigma_1^0$ subset of  $\om^\om$ then the set $L'$ can be chosen to be an (arithmetical) $\Delta_3^0$-subset of the 
Cantor space   $\{0, 1\}^\om$. 
Moreover if $\sigma$ is a winning strategy for Player 1 (respectively, Player 2)  in the game $G(L)$  
then one can construct a w.s. $\sigma'$ for Player 1 (respectively, Player 2)  in the game $G(L')$ such that $\sigma'$ is recursive in $\sigma$.  
And conversely, if     $\sigma$ is a winning strategy for Player 1 (respectively, Player 2)  in the game   $G(L')$ then one                                        
can construct a w.s. $\sigma'$ for Player 1 (respectively, Player 2)  in the game $G(L)$  such that $\sigma'$ is recursive in $\sigma$.  
\end{lem}

\proo  
Let $L \subseteq \om^\om$ be an effective analytic subset of the Baire space, and let $\varphi$ be the mapping from the Baire space $\om^\om$ into the 
Cantor space $\{0, 1\}^\om$ defined by: 
$$\varphi( (n_i) _{i\geq 1}) = (11)^{n'_1}0 (11)^{n'_2}0 \ldots (11)^{n'_i}0 (11)^{n'_{i+1}}0 \ldots$$
\noi where for each integer $i\geq 1$ ~$n_i \in \om$ and  $n'_i=n_i + 1$. 

\hs Notice that $\varphi(\om^\om)=[(11)^+.0]^\om$ is a regular $\om$-language accepted by a deterministic B\"uchi automaton, hence it is an arithmetical 
$\Pi_2^0$-subset of $\{0, 1\}^\om$. 

\hs We now define the set $L'$ as the union of the following sets $D_i$, for $1\leq i \leq 4$: 
\begin{itemize}
\ite $D_1 =\varphi(L)$, 
\ite $D_2 = \{ y \mid  \exists n, k  \geq 0  ~~ y\in  [(11)^+.0]^{2n}.(1)^{2k+1}.0.\{0, 1\}^\om \}$, 
\ite $D_3 = \{ y \mid  \exists n \geq 0 ~~ y \in [(11)^+.0]^{2n+1}.1^\om \}$, 
\ite $D_4 =  \{ y \mid  \exists n \geq 0 ~~ y \in    [(11)^+.0]^{2n+1}.0.\{0, 1\}^\om \}$, 
\end{itemize}

\noi We now explain the meaning of these sets. The first set $D_1$ codes the set $L \subseteq \om^\om$. The other sets 
$D_i$, for $2\leq i \leq 4$ are the results of infinite plays where two players alternatively write letters $0$ or $1$ and the infinite word written 
by the players in $\om$ steps is out of the set $\varphi(\om^\om)$, {\it due to the letters written by Player 2}. 

\hs Notice first that   if  the two players alternatively write letters $0$ or $1$ and the infinite word written 
by the players in $\om$ steps is in the form 
$$\varphi( (n_i) _{i\geq 1}) = (11)^{n'_1}0 (11)^{n'_2}0 \ldots (11)^{n'_i}0 (11)^{n'_{i+1}}0 \ldots$$
\noi then the letters $0$  have been written  alternatively by Player 1 and by Player 2 and  the writing of these letters $0$ determines the integers 
$n'_i$ and therefore also the integers $n_i$. Thus the integers $n_{2i+1}$, $i\geq 0$, have been chosen by Player 1 and the 
integers $n_{2i}$, $i\geq 1$, have been chosen by Player 2. 

\hs We can now see that  $D_2$ is the set of plays where Player 2 write the $(2n+1)$ th letter $0$ while it was  Player 1's turn to do this.   The set $D_3$ 
is the set of plays where Player 2 does not write any letter $0$ for the rest of the play when it is his turn to do this. And the set $D_4$ is the set of plays 
where Player 2 writes a letter $0$ immediately after Player 1 writes a letter $0$, while Player $2$ should then writes a letter $1$ to respect the codes of 
integers given by the function $\varphi$. 

\hs Moreover it is easy to see that the mapping $\varphi$ is a recursive isomorphism between the Baire space $\om^\om$ and 
its image $\varphi(\om^\om) \subseteq \{0, 1\}^\om$ which is an arithmetical 
$\Pi_2^0$-subset of $\{0, 1\}^\om$.  And it is easy to see that $D_2$ and $D_4$ are $\om$-regular (arithmetical) $\Sigma_1^0$-subsets 
of $\{0, 1\}^\om$, and that $D_3$ is an $\om$-regular (arithmetical) $\Sigma_2^0$-subset of $\{0, 1\}^\om$. Therefore this implies the following 
facts: 
\newline (1)  If $L$ is a $\Si_1^1$-subset (respectively,  a $\Delta_1^1$-subset, a $\Sigma_1^0$-subset)   of $\om^\om$ then  $\varphi(L)$ is a $\Si_1^1$-subset 
(respectively,  a $\Delta_1^1$-subset, a $\Delta_3^0$-subset)  of  $\{0, 1\}^\om$. 
\newline  (2)  If $L$ is a $\Si_1^1$-subset (respectively,  a $\Delta_1^1$-subset, a $\Sigma_1^0$-subset)   of $\om^\om$ then  $L'$ is a 
 $\Si_1^1$-subset (respectively,  a $\Delta_1^1$-subset, a $\Delta_3^0$-subset)  of  $\{0, 1\}^\om$.

\hs We now prove that Player 1 (respectively, Player 2)  has a  w.s. in the game $G(L)$ iff Player 1
(respectively, Player 2)  has a  w.s. in the game $G(L')$. 

\hs Assume firstly that Player 1 has a w.s. $F_1$ in the game $G(L)$.  
Consider a play in the game   $G(L')$. If  the two players alternatively write letters $0$ or $1$ and the infinite word written 
by the players in $\om$ steps is in the form 
$$\varphi( (n_i) _{i\geq 1}) = (11)^{n'_1}0 (11)^{n'_2}0 \ldots (11)^{n'_i}0 (11)^{n'_{i+1}}0 \ldots$$
\noi then we have already seen that the integers $n'_{2i+1}$, $i\geq 0$, have been chosen by Player 1 and the 
integers $n'_{2i}$, $i\geq 1$, have been chosen by Player 2, and this is also the case for the corresponding integers  $n_{2i+1}$, $i\geq 0$, and 
$n_{2i}$, $i\geq 1$. Thus the game is like a game where each player writes some integer at each step of the play, and Player 1 can apply  the strategy 
$F_1$ to ensure that $(n_i) _{i\geq 1} \in L$ and this implies that $\varphi( (n_i) _{i\geq 1}) \in \varphi(L) \subseteq L'$, so Player 1 wins the play. 
On the other hand we have seen that if the  two players alternatively write letters $0$ or $1$ and the infinite word $x$ written 
by the players in $\om$ steps is out of the set $\varphi(\om^\om)$, {\it due to the letters written by Player 2}, then  the  $\om$-word $x$ is in 
$D_2 \cup D_3 \cup D_4$, and thus Player 1 wins also the play. Finally this shows that Player 1 has a w. s. in the game  $G(L')$. 

\hs Assume now that Player 2 has a winning strategy $F_2$  in the game $G(L)$. 
 
Consider a play in the game   $G(L')$. If  the two players alternatively write letters $0$ or $1$ and the infinite word written 
by the players in $\om$ steps is in the form 
$$\varphi( (n_i) _{i\geq 1}) = (11)^{n'_1}0 (11)^{n'_2}0 \ldots (11)^{n'_i}0 (11)^{n'_{i+1}}0 \ldots$$
\noi then we have already seen that the integers $n'_{2i+1}$, $i\geq 0$, have been chosen by Player 1 and the 
integers $n'_{2i}$, $i\geq 1$, have been chosen by Player 2, and this is also the case for the corresponding integers  $n_{2i+1}$, $i\geq 0$, and 
$n_{2i}$, $i\geq 1$. Thus the game is like a game where each player writes some integer at each step of the play, and Player 2 can apply  the strategy 
$F_2$ to ensure that $(n_i) _{i\geq 1} \notin L$ and this implies that $\varphi( (n_i) _{i\geq 1}) \notin \varphi(L)$, and also 
$\varphi( (n_i) _{i\geq 1}) \notin L'$ because $L' \cap \varphi(\om^\om) = \varphi(L)$, 
so Player 2 wins the play. 
On the other hand we can easily see that if the  two players alternatively write letters $0$ or $1$ and the infinite word $y$ written 
by the players in $\om$ steps is out of the set $\varphi(\om^\om)$, {\it due to the letters written by Player 1}, then  the  $\om$-word $y$ is not in 
$D_2 \cup D_3 \cup D_4$, and thus $y$ is not in $L'$ and Player 2 wins also the play. 
Finally this shows that Player 2 has a w. s. in the game  $G(L')$. 

\hs Conversely assume now  that Player 1 has a w.s. $F'_1$ in the game $G(L')$.  Consider a play in the game $G(L')$ 
 in which Player 2  does not  make  that the final $\om$-word $x$ written by the two players is in $D_2 \cup D_3 \cup D_4$. 
Then Player 1, following the strategy $F'_1$,   must  write letters so that the final $\om$-word $x$ belongs to $\varphi(\om^\om)$. 
Then the game is reduced to 
the game $G(L)$ in which the two  players alternatively write some integers $n_i$,  $i\geq 1$. But Player 1 wins the game 
and this implies that Player 1 has actually a w.s. in the game $G(L)$. 

\hs Assume now  that Player 2 has a w.s. $F'_2$ in the game $G(L')$.  By a very similar reasoning as in the preceding  case  we can see 
that Player 2 has also a w.s. in the game $G(L)$; details are here left to the reader. 

\hs From the construction of the strategies given in the previous paragraphs, it is now easy to see that 
 if $F$ is a winning strategy for Player 1 (respectively, Player 2)  in the game $G(L)$  
then one can construct a w.s. $F'$ for Player 1 (respectively, Player 2)  in the game $G(L')$ such that $F'$ is recursive in $F$.  
And conversely, if     $F'$ is a winning strategy for Player 1 (respectively, Player 2)  in the game   $G(L')$ then one                                        
can construct a w.s. $F$ for Player 1 (respectively, Player 2)  in the game $G(L)$  such that $F$ is recursive in $F'$.  
\ep 

\hs We can now give the proof of the above Theorem \ref{mod-zfc}.

\hs \noi {\bf Proof of Theorem \ref{mod-zfc}.}  We know from Stern's Theorem \ref{theo-stern} that 
 there exists an effective coanalytic set $L_1 \subseteq \om^\om$ such that 
the Gale-Stewart game $G(L_1)$ is determined if and only if  the ordinal  $\omega_1^{\bf  L}$ is countable. Moreover if 
the game  $G(L_1)$ is determined then  Player 2  has a winning strategy. Then Lemmas \ref{co-an} and \ref{2alph} imply that there 
exists a  effective analytic set $L  \subseteq \{0, 1\}^\om$ such that $G(L)$ is determined if and only if  the ordinal  $\omega_1^{\bf  L}$ is countable.
And moreover if  the game  $G(L)$ is determined then  Player 1  has a winning strategy. 
We can now infer from Propositions \ref{pro-r1c} and \ref{pro-2t} that there 
there  exists a  real-time $1$-counter 
B\"uchi automaton  $\mathcal{A}$,  reading words over a finite alphabet $X$,  and a $2$-tape  B\"uchi automaton  $\mathcal{B}$,   reading words over a finite alphabet $Y$, such that  the  
game $G(L(\mathcal{A}))$ (respectively, $G(L(\mathcal{B}))$) is determined if and only if  $\omega_1^{\bf  L}$ is countable. Moreover 
if  the game $G(L(\mathcal{A}))$ (respectively, $G(L(\mathcal{B}))$) is determined then  Player 1  has a winning strategy. 

\hs   Assume now that $V_1$  is a  model  of  ZFC in which $\omega_1^{\bf  L}$ is countable, i.e. is a model of (ZFC + $\om_1^{\bf L} < \om_1$). 
Then Player 1 has a winning strategy  
in the game $G(L(\mathcal{A}))$. This strategy  is a mapping $F: (X^2)^\star  \ra X$ hence it can be coded in a recursive manner by an infinite word $X_F \in \{0, 1\}^\om$ which 
may be identified with a subset of the set  $\mathbb{N}$ of natural numbers. 
We now claim that this strategy is not constructible, or equivalently that the set $X_F \subseteq \mathbb{N}$ 
does not belong to the class ${\bf L}^{V_1}$ of constructible sets 
in the model $V_1$.  Recall that a  real-time $1$-counter B\"uchi automaton  $\mathcal{A}$ has a finite description to which can be associated, in an effective way,  
 a unique natural number called its index, so we have  a G\"odel numbering of  real-time $1$-counter B\"uchi automata. 
We  denote $\mathcal{A}_z$ the  real time B\"uchi $1$-counter automaton of index $z$ 
reading words over $X$. Then there exists an  integer $z_0$ such that $\mathcal{A}=\mathcal{A}_{z_0}$. 
If $x\in X^\om$ is the $\om$-word written by Player 2 during  a play of the   game  $G(L(\mathcal{A}))$, 
and Player 1 follows a strategy $G$,  the $\om$-word   $(G \star x) \in X^\om$ is defined by 
$(G \star x)(2n)=x(n)$ and $(G \star x)(2n+1)=G((G \star x)[2n])$ for all integers $n\geq 1$ 
so that  $(G \star x)$ is the $\om$-word  composed by the two players during the play. 
We can now easily  see that 
the sentence:
 ``$G$ is a winning strategy for Player 1 in the game $G(L(\mathcal{A}_z))$" can be expressed  by the following   
$\Pi_2^1$-formula  $P(z, G):$ ~~~~~$\fa x \in X^\om ~~[~~(G \star x) \in L(\mathcal{A}_{z}) ~~]$

\hs \noi Recall  that $x\in L(\mathcal{A}_z)$  can be expressed by a $\Si_1^1$-formula (see \cite{Fin-HI}). And $(G \star x) \in L(\mathcal{A}_{z}) $ 
can be expressed by $ \exists y \in X^\om ( y=(G \star x)$ and $y \in L(\mathcal{A}_{z}) )$, which is  also  a  $\Si_1^1$-formula since $(G \star x)$
is recursive in $x$ and $G$. Finally the formula  $P(z,  G)$ is a $\Pi_2^1$-formula (with parameters $z$ and $G$).

\hs 
Towards a contradiction, assume now that the winning strategy $F$ for Player 1 in  the  game $G(L(\mathcal{A}))$ belongs to 
the class ${\bf L}^{V_1}$ of constructible sets in the model $V_1$.  The relation $P_F \subseteq \mathbb{N}$ defined 
by $P_F(z)$ iff $P(z,F)$ is a  $\Pi_2^1(F)$-relation, i.e. a  relation with is $\Pi_2^1$ with  parameter $F$. By Shoenfield's Absoluteness Theorem 
(see \cite[page 490]{Jech}), the relation $P_F \subseteq \mathbb{N}$ would be absolute for the models ${\bf L}^{V_1}$  and $V_1$ 
of   ZFC. This means that the set $\{z \in \mathbb{N} \mid  P_F(z) \}$ would be the same set in the two models 
${\bf L}^{V_1}$  and $V_1$. In particular, the integer  
$(z_0)$ belongs to $P_F$  in the model $V_1$ since $F$ is  a w.s. for   Player 1  in the game $G(L(\mathcal{A}))$. 
 This would imply that $F$ is also a w.s. for   Player 1  in the  game $G(L(\mathcal{A}))$ in the model ${\bf L}^{V_1}$. 
But  ${\bf L}^{V_1}$ is a model of ZFC + V=L so in this model $\om_1^{\bf L} = \om_1$ holds and  the game $G(L(\mathcal{A}))$ is not determined. 
This contradiction shows that the w.s. $F$ is not constructible in $V_1$. On the other hand every set $A \subseteq  \mathbb{N}$ which is 
$\Pi_2^1$ or $\Si_2^1$ is constructible, see \cite[page 491]{Jech}. Thus $X_F$ is neither a $\Pi_2^1$-set nor a $\Si_2^1$-set; in particular,  the 
strategy $F$ is not recursive and not even hyperarithmetical, i.e. not $\Delta_1^1$. 

\hs The case of the game $G(L(\mathcal{B}))$, for the        $2$-tape  B\"uchi automaton  $\mathcal{B}$,   is proved in a similar way. 

\ep 

\begin{Rem}
The  $1$-counter B\"uchi automaton  $\mathcal{A}$ and the  $2$-tape  B\"uchi automaton  $\mathcal{B}$, given by Theorem  \ref{mod-zfc}, 
can be effectively constructed, although the automata might have a great number of states. Indeed the effective coanalytic set $L_1 \subseteq \om^\om$ 
such that 
the Gale-Stewart game $G(L_1)$ is determined if and only if  the ordinal  $\aleph_1^{\bf  L}$ is countable is explicitly given by a formula $\psi$. 
Then the effective analytic set $L  \subseteq \{0, 1\}^\om$ such that $G(L)$ is determined if and only if  the ordinal  $\aleph_1^{\bf  L}$ is countable is also 
given by a $\Sigma_1^1$-formula from which on can construct a B\"uchi Turing machine and thus a 2-counter B\"uchi automaton accepting it. 
The constructions given in the proofs of Propositions \ref{pro-r1c} and \ref{pro-2t} lead then to the effective construction of  $\mathcal{A}$ and $\mathcal{B}$. 
\end{Rem}

\begin{Rem}
In the above proof of  Theorem \ref{mod-zfc} we have not  used any large cardinal axiom  or even the consistency of such an axiom, like the 
axiom of analytic determinacy. 
\end{Rem}

\noi We now prove some lemmas which will be useful later to give some answer to Question 2. 

\begin{lem}\label{L1}
Let $L \subseteq \Si^\om$ be a $\Delta_3^0$-subset of  a  Cantor space, accepted by a 
B\"uchi $2$-counter automaton $\mathcal{A}$ and let $\mathcal{B}$  be the  real time B\"uchi $1$-counter automaton which can be 
effectively constructed from $\mathcal{A}$ by Proposition \ref{pro-r1c}. Then $L(\mathcal{B})$ is also a $\Delta_3^0$-subset of a  Cantor space
$Y^\om$ for some finite alphabet $Y$ containing $\Sigma$. 
\end{lem}

\proo  We refer now to the proofs of  Proposition 
 \ref{the1}  and Theorems \ref{the2} and  \ref{the3} in \cite{Fin13-JSL}, and we  use here the same notations as in \cite{Fin13-JSL}. 

\hs In the proof of Proposition \ref{the1} it is firstly proved that, from a B\"uchi $2$-counter automaton $\mathcal{A}$ accepting $L$, one can construct 
a real time B\"uchi $8$-counter automaton $\mathcal{A}_3$ accepting $\theta_S(L) \cup  L'$, where $\theta_S: \Sio \ra (\Sigma \cup \{E\})^\om$ is a  
function defined, for all  $x \in \Sio$, by: 
$$\theta_S(x)=x(1).E^{S}.x(2).E^{S^2}.x(3).E^{S^3}.x(4) \ldots 
x(n).E^{S^n}.x(n+1).E^{S^{n+1}} \ldots $$
\noi It is easy to see that $\theta_S$ is a recursive homeomorphism from $\Sio$ onto the image $\theta_S(\Sio)$ which is a closed subset of 
the Cantor space      $(\Sigma \cup \{E\})^\om$. It is then easy to se that  if $L$ is a $\Delta_3^0$-subset of  $\Sio$ then $\theta_S(L)$ is also 
a $\Delta_3^0$-subset $(\Sigma \cup \{E\})^\om$. Moreover the $\om$-language $L'$  is defined as the set of $\om$-words 
$y \in (\Sigma \cup \{E\})^\om$ for which  there is an integer $n\geq 1$ such that 
$y[2n-1]\in {\rm Pref}(\theta_S(\Sio))$ and $y[2n]\notin {\rm Pref}(\theta_S(\Sio))$. Then it is easy to see that $L'$ is an arithmetical 
$\Sigma_1^0$-subset of  $(\Sigma \cup \{E\})^\om$, and thus the union $\theta_S(L) \cup  L'$ is a $\Delta_3^0$-set as the union of two 
$\Delta_3^0$-sets. 

Recall also that Player 1 (respectively, Player 2)  has a  w.s. in the game  $G(L)$ iff Player 1
(respectively, Player 2)  has a  w.s. in the game   $G( \theta_S(L) \cup  L' )$. 

\hs In a second step, in the proof of Theorem \ref{the2}, it is proved that, from a real time B\"uchi $8$-counter automaton $\mathcal{A}$ accepting an 
$\om$-language $L(\mathcal{A}) \subseteq \Gao$, where $\Ga$ is a finite alphabet, one can construct a B\"uchi $1$-counter automaton $\mathcal{A}_4$
accepting the $\om$-language 
$$\mathcal{L}~ ~   =  ~ ~  h( L(\mathcal{A}) ) ~ ~  \cup ~ ~   [ h(\Ga^{\om})^- \cap H ] ~ ~   \cup~ ~    V.C^\om ~ ~  \cup ~ ~  U.(\Ga_1)^\om$$
Moreover it is proved that Player 1 (respectively, Player 2)  has a  w.s. in the game  $G(L(\mathcal{A}))$ iff Player 1
(respectively, Player 2)  has a  w.s. in the game   $G(\mathcal{L})$.

On the other hand the mapping $h$ is a recursive homeomorphism from $\Ga^{\om}$ onto its image $h(\Ga^{\om}) \subseteq (\Ga_1)^{\om}$
where $\Ga_1$ is the finite alphabet $\Ga \cup \{A, B, C\}$ and $A, B, C$, are additional letters not in $\Ga$. It is then easy to see that 
if $L(\mathcal{A}) \subseteq \Gao$ is a $\Delta_3^0$-set then $h( L(\mathcal{A}) )$ is a $\Delta_3^0$-subset of $(\Ga_1)^{\om}$. 
On the other hand the $\om$-language $H$ is accepted by a deterministic B\"uchi automaton and hence it is an arithmetical $\Pi_2^0$-set, 
see \cite{PerrinPin,LescowThomas}. Thus  $[ h(\Ga^{\om})^- \cap H ] $ is also a $\Pi_2^0$-set since it is the intersection of a $\Sigma_1^0$-set and of a 
$\Pi_2^0$-set. Moreover it is easy to see that $V.C^\om$ is a $\Si_2^0$-set since it is accepted by a deterministic automaton with co-B\"uchi 
acceptance condition,  and that $U.(\Ga_1)^\om$ is a $\Sigma_1^0$-subset of  $(\Ga_1)^\om$ since $U$ is regular and hence recursive. Finally 
this shows that if $L(\mathcal{A}) \subseteq \Gao$ is a $\Delta_3^0$-set then $\mathcal{L}$ is a $\Delta_3^0$-subset of $(\Ga_1)^{\om}$. 

\hs In a third step, in the proof of Theorem \ref{the3}, it is proved that, from the  
B\"uchi $1$-counter automaton $\mathcal{A}_4$ accepting the $\om$-language $\mathcal{L}$, one can construct a 
{\it real time} B\"uchi $1$-counter automaton $\mathcal{B}''$ accepting the 
 $\om$-language $\phi_K(L(\mathcal{A}_4)) \cup L''$.  It is easy to see, as in the  above first step, that if 
$\mathcal{L}=L(\mathcal{A}_4)$ is a $\Delta_3^0$-subset of $(\Ga_1)^{\om}$, then the  $\om$-language $\phi_K(L(\mathcal{A}_4)) \cup L''$ is 
also a $\Delta_3^0$-subset of  $(\Ga_1 \cup \{F\})^{\om}$. 
Moreover Player 1 (respectively, Player 2)  has a  w.s. in the game  $G(\mathcal{L})$ iff Player 1
(respectively, Player 2)  has a  w.s. in the game   $G(\phi_K(\mathcal{L}) \cup L'')$.
\ep 

\begin{lem}\label{L2}
Let $\mathcal{A}$ be a real time B\"uchi $1$-counter automaton accepting a    $\Delta_3^0$-set   $L  \subseteq \Si^\om$ 
and let $\mathcal{B}$  be the  $2$-tape B\"uchi automaton which can be 
effectively constructed from $\mathcal{A}$ by Proposition \ref{pro-2t}. Then $L(\mathcal{B})$ is a 
$\Delta_3^0$-subset of the   Cantor space $(\Si\cup\{0,A\})^\om \times \Ga^\om$, where $0, A$ are additional letters not in $\Si$ and 
$\Ga=\{0, A\}$. 
\end{lem}

\proo We refer now to the proof of  the above Theorem \ref{the4}  and we  use here the same notations. 
We showed  above that, from a a real-time 
$1$-counter B\"uchi automaton $\mathcal{A}$ accepting an  $\om$-language $L= L(\mathcal{A})\subseteq \Si^\om$, we can effectively construct a 
$2$-tape B\"uchi automaton $\mathcal{D}$ accepting the  infinitary rational relation $\mathcal{L}\subseteq (\Si\cup\{0, A\})^\om\times \Ga^\om$, where 
the letters  $0, A$ are  not in $\Si$ and $\Ga=\{0, A\}$, and 
$$\mathcal{L}~ ~   =  ~ ~  L(\mathcal{B}')~ ~  \cup ~ ~   L(\mathcal{C}) ~ ~   \cup~ ~ L(\mathcal{C}')  $$
\noi where 
$$L(\mathcal{B}')= [h(L(\mathcal{A}))  \times \{\alpha\}] \cup  [( h(\Si^{\om}) \times \{\alpha\})^- \cap H \times H'] $$
$$L(\mathcal{C})= [  V.0^\om \times {\rm Cl}(H') ] ~ \cup ~ [ {\rm Cl}(H) \times V'.0^\om ]$$
$$L(\mathcal{C}')=  U.[ (\Si\cup\{0, A\})^\om \times \Gao ] $$

\noi We now assume that $L=L(\mathcal{A})$  is a $\Delta_3^0$-subset of  $\Si^\om$. 

\hs It is easy to see that  the mapping $h$ is a recursive homeomorphism 
from $\Si^{\om}$ onto its image $h(\Si^{\om}) \subseteq (\Si\cup\{0, A\})^{\om}$. Moreover $\alpha$ is recursive and $\{\alpha\}$ is a 
$\Pi_1^0$-subset of  $\Ga^\om$. Therefore $h(L(\mathcal{A}))  \times \{\alpha\}$ is a $\Delta_3^0$-subset of  $(\Si\cup\{0, A\})^\om\times \Ga^\om$. 
On the other hand $( h(\Si^{\om}) \times \{\alpha\})$ is a  $\Pi_1^0$-set, 
and so $(h(\Si^{\om}) \times \{\alpha\})^-$ is a $\Si_1^0$-subset of  $(\Si\cup\{0, A\})^\om\times \Ga^\om$. And it is easy to see that $H$ and $H'$ are 
accepted by deterministic B\"uchi automata and thus are (arithmetical) $\Pi_2^0$-sets. Thus $[( h(\Si^{\om}) \times \{\alpha\})^- \cap H \times H'] $ is also 
a $\Pi_2^0$-set and finally this shows that $L(\mathcal{B}')$ is a 	$\Delta_3^0$-set.  

\hs The $\om$-languages $H$ and $H'$ being $\om$-regular, their closures {\rm Cl}(H) and {\rm Cl}(H') are closed and $\om$-regular and thus they 
are  (arithmetical) $\Pi_1^0$-sets (see \cite{PerrinPin,LescowThomas}) . On the other hand the finitary languages $V$ and $V'$ are regular thus 
$V.0^\om$ and $V'.0^\om$ are  (arithmetical) $\Si_2^0$-sets.  This implies  that 
$L(\mathcal{C})= [  V.0^\om \times {\rm Cl}(H') ] ~ \cup ~ [ {\rm Cl}(H) \times V'.0^\om ]$ is also a $\Delta_3^0$-set. 

\hs The $\om$-language $L(\mathcal{C}')$ is an open $\om$-regular set since the finitary language $U$ is regular. Thus  $L(\mathcal{C}')$ is also 
an (arithmetical) $\Si_1^0$-set. 

\hs Finally the $\om$-language $\mathcal{L}$ is the union of three $\Delta_3^0$-sets and thus it is also a $\Delta_3^0$-set. 
\ep 

\hs We can now state the following result which gives an answer to Question 2. 

\begin{theorem}\label{non-hyper}
There  exist a  real-time $1$-counter 
B\"uchi automaton  $\mathcal{A}$ and a $2$-tape  B\"uchi automaton  $\mathcal{B}$  such that the $\om$-language $L(\mathcal{A})$ and the infinitary 
rational relation $L(\mathcal{B})$ are arithmetical $\Delta_3^0$-sets and such that Player 2 has a winning strategy in the games $G(L(\mathcal{A}))$ 
and $G(L(\mathcal{B}))$ but has no hyperarithmetical winning strategies in these games. 
\end{theorem}

\proo   It is proved in   \cite[Theorem 3]{Blass} that there exists an arithmetical $\Sigma_1^0$-set $L \subseteq \om^\om$ such that Player 2 
has a winning strategy in the game $G(L)$ but has no hyperarithmetical winning strategies in this game. 
Using Lemmas \ref{2alph},  \ref{L1},  \ref{L2},  we see  that one can construct 
a real-time $1$-counter 
B\"uchi automaton  $\mathcal{A}$ and a $2$-tape  B\"uchi automaton  $\mathcal{B}$  such that the $\om$-language $L(\mathcal{A})$ and the infinitary 
rational relation $L(\mathcal{B})$ are arithmetical $\Delta_3^0$-sets and such that  Player 2 
has a winning strategy in the games $G(L(\mathcal{A}))$ 
and $G(L(\mathcal{B}))$. 

Moreover, by Propositions \ref{pro-recws} and \ref{pro-2t},  if $F$ was an  hyperarithmetical winning strategy 
 for Player 2 in the game  $G(L(\mathcal{A}))$ 
or $G(L(\mathcal{B}))$ then there would exist a winning strategy  $T$  for Player 2 in the game $G(L)$ which would be  recursive in $F$ and thus 
also   hyperarithmetical. This implies that 
$F$ can not be hyperarithmetical since Player 2 has no  hyperarithmetical winning strategies in the game $G(L)$. 
\ep 

\hs The above negative results given by Theorems \ref{mod-zfc} and  \ref{non-hyper} show that one cannot effectively construct winning strategies 
in Gale-Stewart games with winning sets accepted by $1$-counter B\"uchi automata or $2$-tape  B\"uchi automata. 
We are going to see that, even when we know that the games are determined,   one cannot  determine the winner of such games. 

\begin{theorem}
There exists a recursive sequence of  real time $1$-counter B\"uchi automata $\mathcal{A}_n$,  (respectively, of $2$-tape  B\"uchi automata $\mathcal{B}_n$), 
$n\geq 1$,  such that all games   $G(L(\mathcal{A}_n))$  (respectively, $G(L(\mathcal{B}_n))$) are determined. But it is $\Pi_2^1$-complete (hence 
highly undecidable) to determine whether Player 1 has a winning strategy in the game $G(L(\mathcal{A}_n))$  (respectively, $G(L(\mathcal{B}_n))$). 
\end{theorem}

\proo  We first define the following operation on $\om$-languages. For $x, x' \in \Sio$ the $\om$-word $x \otimes x'$ is defined by : 
for every integer $n\geq 1$ ~$(x \otimes x')(2n -1)=x(n)$ and $(x \otimes x')(2n)=x'(n)$. 
For two $\om$-languages $L, L' \subseteq \Sio$, the $\om$-language $L \otimes L' $ is defined by $L \otimes L' = \{ x \otimes x' \mid x\in L \mbox{ and } 
x'\in L' \}$. 
 Let   now $\Sigma=\{0, 1\}$ and let  $T_n$ be the B\"uchi Turing machine of index $n$ reading $\om$-words over the alphabet $\Sigma$. 
Let also $\mathcal{T}_n$ be a B\"uchi  Turing machine constructed from $T_n$ such that $L(\mathcal{T}_n) = \Sio \otimes L(T_n)$. 
Notice that $\mathcal{T}_n$ can easily be constructed in a recursive manner from  $T_n$, and that on can also construct some B\"uchi 2-counter automata 
$\mathcal{C}_n$ such that $L(\mathcal{T}_n)=L(\mathcal{C}_n)$. 
 
\hs Consider now the game $G(L(\mathcal{C}_n))$. It is easy to see that this game is always determined. 
Indeed if  $L(T_n)=\Sio$ then Player 1 always wins the 
play so Player 1 has an obvious winning strategy. And if $L(T_n) \neq \Sio$ then Player 2 can win by playing an $\om$-word not in $L(T_n)$ so that 
the final $\om$-word written by the two players will be outside  $L(\mathcal{C}_n) = \Sio \otimes L(T_n)$.  
Recall now that  Castro and Cucker  proved in \cite{cc} that it is  $\Pi_2^1$-complete (hence 
highly undecidable) to determine whether $L(T_n) = \Sio$. Thus it is $\Pi_2^1$-complete (hence 
highly undecidable) to determine whether Player 1 has a winning strategy in the game $G(L(\mathcal{C}_n))$.

\hs Using the constructions we made in the proofs of  Theorems \ref{the3} and \ref{the4} and Propositions  \ref{pro-r1c} and  \ref{pro-2t}, 
we can effectively construct  from $\mathcal{C}_n$
a real time B\"uchi $1$-counter automaton $\mathcal{A}_n$ and a 2-tape B\"uchi   automaton $\mathcal{B}_n$ 
such that Player 1 (respectively, Player 2)  has a w.s. in the game  $G(L(\mathcal{C}_n))$ iff 
Player 1 (respectively, Player 2)  has a w.s. in the game  $G(L(\mathcal{A}_n))$ iff 
Player 1 (respectively, Player 2)  has a w.s. in the game  $G(L(\mathcal{B}_n))$. This implies that it is $\Pi_2^1$-complete (hence 
highly undecidable) to determine whether Player 1 has a winning strategy in the game $G(L(\mathcal{A}_n))$  (respectively, $G(L(\mathcal{B}_n))$). 
\ep 

\hs We now consider the strength of determinacy of a game $G(L(\mathcal{A}))$, where $\mathcal{A}$ is a B\"uchi $1$-counter automaton
or a  2-tape B\"uchi   automaton.  We first recall that there exists some effective analytic set
 $L_\sharp \subseteq \{0, 1\}^\om$ such that the determinacy of the 
game $G(L_\sharp)$ is equivalent to the effective analytic determinacy, i.e. to the determinacy of all effective analytic Gale-Stewart games: a first example 
was given by Harrington in \cite{Harrington}, Stern gave another one in \cite{Stern82}. 
We can now infer from this result a similar one for games specified by automata. 

\begin{theorem}
There exists a real time 1-counter B\"uchi automaton $A_\sharp$ (respectively,  a 2-tape B\"uchi   automaton  $B_\sharp$) 
such  that the game $G(A_\sharp)$ (respectively,   the game $G(B_\sharp)$)     is determined iff the effective analytic determinacy 
holds iff all 1-counter games are determined   iff  all games specified by 2-tape B\"uchi   automata are determined. 
\end{theorem}

\proo  The effective analytic set $L_\sharp \subseteq \{0, 1\}^\om$ is defined by a $\Sigma_1^1$-formula from which one can construct a 
B\"uchi Turing machine and a 2-counter B\"uchi automaton $\mathcal{C}_\sharp$ accepting it. 
 Using the constructions we made in the proofs of  Theorems \ref{the3} and \ref{the4}, 
we can effectively construct  from $\mathcal{C}_\sharp$
a real time B\"uchi $1$-counter automaton $\mathcal{A}_\sharp$ and a 2-tape B\"uchi   automaton $\mathcal{B}_\sharp$ 
such that  the game  $G(L(\mathcal{C}_\sharp))$ is determined iff 
 the game  $G(L(\mathcal{A}_\sharp))$ is determined  iff 
 the game  $G(L(\mathcal{B}_\sharp))$ is determined. 
\ep 

\hs This shows that there exists a  real time 1-counter B\"uchi automaton $A_\sharp$ (respectively,  a 2-tape B\"uchi   automaton  $B_\sharp$) 
such that the determinacy strength of the game $G(L(\mathcal{A}_\sharp))$ (respectively,  $G(L(\mathcal{B}_\sharp))$) is the strongest possible. 
Then the following question naturally arises. 

 \hs \noi {\bf Question 3.} Are there many different strengths of determinacy for games specified by 1-counter B\"uchi automata 
(respectively,  by  2-tape B\"uchi   automata) ? 

\hs We now give a positive answer to this question, stating  the following result. 
Notice that below  $Det(G(L))$ means ``the game $G(L)$ is determined". We  recall that $\om_1^{\rm{CK}}$ 
is the Church-Kleene ordinal, which is the first non-recursive ordinal. 

\begin{theorem}\label{transfinite}
There is a  transfinite sequence of real-time $1$-counter  B\"uchi automata  ($\mathcal{A}_\alpha$)$_{\alpha<\om_1^{\rm{CK}}}$,  (respectively,  
of  2-tape B\"uchi   automata ($\mathcal{B}_\alpha$)$_{\alpha<\om_1^{\rm{CK}}}$), 
indexed by recursive ordinals, s.t.:
 $$ \forall  \alpha < \beta < \om_1^{\rm{CK}}~~ [ ~   Det(G(L(\mathcal{A}_\beta)))   \Longrightarrow  Det(G(L(\mathcal{A}_\alpha)))  ~] $$
 $$ \forall  \alpha < \beta < \om_1^{\rm{CK}}~~ [ ~   Det(G(L(\mathcal{B}_\beta)))   \Longrightarrow  Det(G(L(\mathcal{B}_\alpha)))  ~] $$

but the converse is not true: 

\hs For each recursive ordinal $\alpha$ there is a model {\bf  V}$_\alpha$ of    ZFC  such that in this model 
the game $G(L(\mathcal{A}_\beta))$ (respectively, $G(L(\mathcal{B}_\beta))$)  is determined iff  $\beta  < \alpha$. 
\end{theorem}

\proo  It follows from Stern's Theorem \ref{theo-stern} and from Lemmas  \ref{co-an} and  \ref{2alph} that for each recursive ordinal $\xi$ there exists 
an effective analytic set $L_\xi \subseteq \{0, 1\}^\om$ such that the game $G(L_\xi)$ is determined if and only if  
the ordinal  $\aleph_\xi^{\bf  L}$ is countable.  Notice that  each set $L_\xi$ is accepted by a B\"uchi Turing machine $T_\xi$ 
and by  a 2-counter B\"uchi automaton $\mathcal{C}_\xi$. 

 Using the constructions we made in the proofs of  Theorems \ref{the3} and \ref{the4} and Propositions  \ref{pro-r1c} and  \ref{pro-2t}, 
we can  construct  from $\mathcal{C}_\xi$
a real time B\"uchi $1$-counter automaton $\mathcal{A}'_\xi$ and a 2-tape B\"uchi   automaton $\mathcal{B}'_\xi$ 
such that Player 1 (respectively, Player 2)  has a w.s. in the game  $G(L(\mathcal{C}_\xi))$ iff 
Player 1 (respectively, Player 2)  has a w.s. in the game  $G(L(\mathcal{A}'_\xi))$ iff 
Player 1 (respectively, Player 2)  has a w.s. in the game  $G(L(\mathcal{B}'_\xi))$.  Thus  the game $G(L(\mathcal{A}'_\xi))$ is determined if and only if  
 the game $G(L(\mathcal{B}'_\xi))$ is determined if and only if the ordinal  $\aleph_\xi^{\bf  L}$ is countable.  
We set  $\mathcal{A}_\xi=\mathcal{A}'_{\xi +1}$ and $\mathcal{B}_\xi=\mathcal{B}'_{\xi +1}$. 

The first part of the theorem follows easily from 
the obvious implication $[  \aleph_\xi^{\bf  L} \mbox{ is countable } ]$ $\Longrightarrow  [\aleph_\alpha^{\bf  L} 
\mbox{ is countable, for all  ordinals } \alpha < \xi ].$

Let now $\alpha$ be a recursive ordinal  and  {\bf  V} be a model of  ZFC + V=L.  The cardinal $\aleph_{\alpha +1}$ in {\bf  V} 
 is a successor cardinal  hence also a regular cardinal (the reader may find these notions in any textbook of set theory like \cite{Kunen80} or \cite{Jech}). 
One can then construct  from the model {\bf  V},  using a forcing method due to L\'evy,  a generic extension {\bf  V}$_\alpha$ of  {\bf  V} which is another 
model of   ZFC in which the cardinal $\aleph_{\alpha +1}$ has been ``collapsed"  in such a way that in the new model $\aleph_{\alpha +1}$ becomes 
$\omega_1^{{\bf  V}_\alpha}$.  Notice that the two models have the same ordinals, and the above sentence means that 
the ordinal of  {\bf  V} which plays the role of  
$\aleph_{\alpha +1}$ in {\bf  V} plays the role of the cardinal $\aleph_{1}$ in {\bf  V}$_\alpha$ (we refer the  
reader to \cite[page 231]{Kunen80} for more details about Lévy's forcing). 

Another crucial point here is that the two models {\bf  V} and {\bf  V}$_\alpha$ have the same constructible sets (this is always true for generic extensions 
obtained by the method of forcing), i.e. 
${\bf  L}^{{\bf  V}}={\bf  L}^{{\bf  V}_\alpha}$.     Notice also 
 that $\aleph_{\alpha +1}^{{\bf  L}}=\aleph_{\alpha +1}$ since {\bf  V} is  a model of   ZFC + V=L. For a recursive ordinal $\beta$, we 
 have now the following equivalences: 

\hs \noindent  [ $\aleph_{\beta +1}^{{\bf  L}}$ is countable in {\bf  V}$_\alpha$ ]
 $\Longleftrightarrow [ \aleph_{\beta +1}^{{\bf  L}} <  \omega_1^{{\bf  V}_\alpha} = \aleph_{\alpha +1}^{{\bf  L}}$ ]
  $\Longleftrightarrow  \beta +1  <   \alpha +1$
  $\Longleftrightarrow  \beta  <   \alpha $

\hs And thus $G(L(\mathcal{A}_\beta))$ (respectively,  $G(L(\mathcal{B}_\beta))$      is determined in the model   {\bf  V}$_\alpha$
   if and only if  $\beta  <   \alpha$. 
\ep 

\begin{Rem}
We can add the   real time 1-counter B\"uchi automaton $A_\sharp$ and  the  2-tape B\"uchi   automaton  $B_\sharp$ to the sequences given 
by Theorem \ref{transfinite}.  The determinacy of  $G(L(\mathcal{A}_\sharp))$ (respectively,  $G(L(\mathcal{B}_\sharp))$)  implies the determinacy of 
all games $G(L(\mathcal{A}_\alpha))$ (respectively,  $G(L(\mathcal{B}_\alpha))$, $\alpha<\om_1^{\rm{CK}}$, but the converse is not true. 
Then we get a transfinite sequence of  real time 1-counter B\"uchi automata (respectively, of 2-tape B\"uchi   automata) of length 
$\om_1^{\rm{CK}} +1$.  
\end{Rem}

\begin{Rem}
One can actually see from \cite{McAloon79} that  the situation is even more complicated.  Indeed Mc Aloon proved 
that there exists some analytic game whose determinacy is equivalent to the fact that the first inaccessible cardinal in the constructible universe 
{\bf  L} of a model  {\bf  V} of    ZFC    is countable in {\bf  V}.  And this property implies that  $\aleph_{\alpha}^{{\bf  L}}$, for a recursive ordinal 
$\alpha$, is countable in {\bf  V}, but does not imply the existence of $0^\sharp$. 
We refer the interested reader to \cite{Jech} for  the notion of inaccessible cardinals and of other  large cardinals, and to 
  \cite{McAloon79} for more results of this kind. 
\end{Rem}

\section{Wadge games between  2-tape automata}

The now called  Wadge games have been firstly considered by Wadge to study the notion of reduction of  Borel sets by continuous functions. 
We firstly recall the notion of Wadge reducibility; notice that we 
 give the definition in the case of $\om$-languages over {\it finite} alphabets since we have only to 
consider  this case in the sequel. 

\begin{defi}[Wadge \cite{Wadge83}] Let $X$, $Y$ be two finite alphabets. 
For $L\subseteq X^\om$ and $L'\subseteq Y^\om$, $L$ is said to be Wadge reducible to $L'$
($L\leq _W L')$ iff there exists a continuous function $f: X^\om \ra Y^\om$, such that
$L=f^{-1}(L')$.
 $L$ and $L'$ are Wadge equivalent iff $L\leq _W L'$ and $L'\leq _W L$. 
This will be denoted by $L\equiv_W L'$. And we shall say that 
$L<_W L'$ iff $L\leq _W L'$ but not $L'\leq _W L$.

\noi
 The relation $\leq _W $  is reflexive and transitive,
 and $\equiv_W $ is an equivalence relation.
\nl The {\it equivalence classes} of $\equiv_W $ are called {\it Wadge degrees}. 
\end{defi}

We now recall the definition of Wadge games. 

\begin{defi}[Wadge \cite{Wadge83}]  Let 
$L\subseteq X^\om$ and $L'\subseteq Y^\om$. 
The Wadge game  $W(L, L')$ is a game with perfect information between two players,
Player 1 who is in charge of $L$ and Player 2 who is in charge of $L'$.
 Player 1 first writes a letter $a_1\in X$, then Player 2 writes a letter
$b_1\in Y$, then Player 1 writes a letter $a_2\in  X$, and so on. 
 The two players alternatively write letters $a_n$ of $X$ for Player 1 and $b_n$ of $Y$
for Player 2.
 After $\om$ steps,  Player 1 has written an $\om$-word $a\in X^\om$ and  Player 2
has written an $\om$-word $b\in Y^\om$.
 Player 2 is allowed to skip, even infinitely often, provided he really writes an
$\om$-word in  $\om$ steps.
Player 2 wins the play iff [$a\in L \lra b\in L'$], i.e. iff: 
~~~~  [($a\in L ~{\rm and} ~ b\in L'$)~ {\rm or} ~ 
($a\notin L ~{\rm and}~ b\notin L'~{\rm and} ~ b~{\rm is~infinite}  $)].
\end{defi}

\noi
Recall that a strategy for Player 1 is a function 
$\sigma :(Y\cup \{s\})^\star\ra X$.
And a strategy for Player 2 is a function $f:X^+\ra Y\cup\{ s\}$.
The strategy  $\sigma$ is a winning strategy  for Player 1 iff she always wins a play when
 she uses the strategy $\sigma$, i.e. when the  $n^{th}$  letter she writes is given
by $a_n=\sigma (b_1\ldots b_{n-1})$, where $b_i$ is the letter written by Player 2 
at step $i$ and $b_i=s$ if Player 2 skips at step $i$.
 A winning strategy for Player 2 is defined in a similar manner.

\hs The game $W(L, L')$ is said to be determined if one of the two players has a winning strategy. 
\noi In the sequel we shall denote {\bf W-Det}($\mathcal{C}$), where $\mathcal{C}$ is a class of $\om$-languages, 
the sentence: ``All Wadge games $W(L, L')$,  where $L\subseteq X^\om$ and  $L'\subseteq Y^\om$ are $\om$-languages 
in the class $\mathcal{C}$, are determined". 

\hs Recall that the determinacy of  Borel Gale-Stewart games implies easily the determinacy of Wadge games $W(L, L')$,  
where $L\subseteq X^\om$ and  $L'\subseteq Y^\om$ are Borel $\om$-languages. Thus it follows from Martin's Theorem 
that these Wadge games are determined.  We also recall that the determinacy of  effective analytic Gale-Stewart games is equivalent to 
the determinacy of  effective analytic  Wadge  games,  i.e. {\bf Det}($\Si_1^1$)   
$\Longleftrightarrow$ {\bf W-Det}($\Si_1^1$), see  \cite{Louveau-Saint-Raymond}.

\hs  The close relationship between Wadge reducibility
 and Wadge games is given by the following theorem.  

\begin{theorem} [Wadge] Let $L\subseteq X^\om$ and $L'\subseteq Y^\om$  where
$X$ and $Y$ are finite  alphabets. Then  $L\leq_W L'$ if and only if  Player 2 has a 
winning strategy  in the Wadge game $W(L, L')$.
\end{theorem}

The Wadge hierarchy $WH$ is the class of Borel subsets of a set  $X^\om$, where  $X$ is a finite set,
equipped with $\leq _W $ and with $\equiv_W $. Using Wadge games, Wadge proved that, up to the complement and $\equiv _W$, 
it is a well ordered hierarchy which 
provides a  great refinement of the Borel hierarchy. 

\begin{theorem} [Wadge]\label{wh}
The class of Borel subsets of $X^\om$,
 for  a finite alphabet $X$,  equipped with $\leq _W $,  is a well ordered hierarchy.
 There is an ordinal $|WH|$, called the length of the hierarchy, and a map
$d_W^0$ from $WH$ onto $|WH|-\{0\}$, such that for all $L, L' \subseteq X^\om$:
\nl $d_W^0 L < d_W^0 L' \lra L<_W L' $  and 
\nl $d_W^0 L = d_W^0 L' \lra [ L\equiv_W L' $ or $L\equiv_W L'^-]$.
\end{theorem}

 We proved in \cite{Fin13-JSL} the following result on  the determinacy of Wadge games between two players in charge of 
$\om$-languages of one-counter automata.

\begin{theorem}\label{thew}
{\bf Det}($\Si_1^1$)   
$\Longleftrightarrow$ {\bf W-Det}({\bf r}-${\bf BCL}(1)_\om$). 
\end{theorem}

Using this result we are now going to prove the following one on determinacy of  Wadge games between two players in charge of 
$\om$-languages accepted by  2-tape B\"uchi   automata. 

\begin{theorem}\label{thew-rat}
{\bf Det}($\Si_1^1$)  $\Longleftrightarrow$   {\bf W-Det}(${\bf RAT}_\om$). 
\end{theorem}

In order to prove this theorem, we first  recall the notion of   operation of sum of 
sets of infinite words which has as  
counterpart the ordinal
addition  over Wadge degrees, and which will useful  later.

\begin{defi}[Wadge]
Assume that $X\subseteq Y$ are two finite alphabets,
  $Y-X$ containing at least two elements, and that
$\{X_+, X_-\}$ is a partition of $Y-X$ in two non empty sets.
 Let $L \subseteq X^{\om}$ and $L' \subseteq Y^{\om}$, then
 $$L' + L =_{df} L\cup \{ u.a.\beta  ~\mid  ~ u\in X^\star , ~(a\in X_+
~and ~\beta \in L' )~
or ~(a\in X_- ~and ~\beta \in L'^- )\}$$
\end{defi}

Notice that a  player in charge of a set $L'+L$ in a Wadge game is like a player in charge of the set $L$ but who 
can, at any step of the play,    erase  his previous play and choose to be this time in charge of  $L'$ or of $L'^-$. 
But he can do this only one time during a play. This property will be used below. 

We now recall  the following lemma, proved in \cite{Fin13-JSL}. 

\begin{lem}\label{w1}
Let $L \subseteq \Sio$ be an analytic but non Borel set. Then it holds that $L \equiv_W \emptyset + L$. 
\end{lem}

\noi Notice that in this lemma, $\emptyset$ is viewed as the empty set over an alphabet $\Gamma$ such that 
$\Si \subseteq \Ga$ and cardinal ($\Ga - \Si$) $\geq 2$.  Recall also that the emptyset and  the whole set $\Gao$ are located 
 at the first level of the Wadge hierarchy and that their Wadge degree is equal to 1. 

\hs \noi {\bf proof of Theorem \ref{thew-rat}.}

The implication {\bf  Det}($\Si_1^1$)  $\Longrightarrow${\bf W-Det}(${\bf RAT}_\om$) is obvious since  
{\bf Det}($\Si_1^1$)   is known to be equivalent to  {\bf W-Det}($\Si_1^1$) and 
${\bf RAT}_\om$ $\subseteq \Si_1^1$. 

 To prove the reverse implication, we assume that {\bf W-Det}(${\bf RAT}_\om$) holds  and we are going to show that every Wadge game 
$W(L, L')$,  where $L \subseteq (\Si_1)^\om$ and $L' \subseteq (\Si_2)^\om$ are $\om$-languages in the class {\bf r}-${\bf BCL}(1)_\om$, 
 is determined. Then this will imply that {\bf  Det}($\Si_1^1$)  holds by Theorem \ref{thew}. 
Notice that if the two $\om$-languages are Borel we already know that the game $W(L, L')$ is determined; thus we have only to consider the case where 
at least one of these languages is non-Borel. 

 We  now assume that the letters $0$ and $A$ do not belong to the alphabets $\Si_1$ and $\Si_2$, and recall that we have used in the proof  of Theorem 
\ref{the4}  a mapping $h_1:  (\Si_1)^\om \ra (\Si_1 \cup \{ 0, A \})^\om$  to code $\om$-words over $\Si_1$ by  $\om$-words over $\Si_1 \cup\{ 0, A \}$; 
and we can define similarly  $h_2:  (\Si_2)^\om \ra (\Si_2 \cup\{ 0, A \})^\om$. Recall also that we have defined an $\om$-word 
$\alpha \in \{0, A\}^\om=\Ga^\om$. 

 It follows from   Lemmas \ref{R1} and  \ref{complement} that one can effectively construct,  from 
real-time   B\"uchi $1$-counter automata  $\mathcal{A}_1$  and  $\mathcal{A}_2$ accepting $L$ and $L'$, some 
$2$-tape B\"uchi  automata $\mathcal{B}_1$  and  $\mathcal{B}_2$ accepting the $\om$-languages 

$$\mathcal{L}_1 =  [ h_1(L) \times \{\alpha\} ]  \cup [ h_1(\Si_1^{\om}) \times \{\alpha\} ]^-$$
and 
$$\mathcal{L}_2 =  [ h_2(L') \times \{\alpha\}  ]  \cup [ h_2(\Si_2^{\om}) \times \{\alpha\} ]^-$$

Then the Wadge game 
$W(\mathcal{L}_1, \mathcal{L}_2)$ is determined. We consider now the two following cases: 

\hs \noi  {\bf First case.} Player 2 has a w.s. in the game $W(\mathcal{L}_1, \mathcal{L}_2)$. 

If $L'$ is Borel then $h_2(L') \times \{\alpha\}$ is easily seen to be Borel and then $\mathcal{L}_2$ is also Borel since $h_2(\Si_2^{\om}) \times \{\alpha\}$
is a closed set and hence $[ h_2(\Si_2^{\om}) \times \{\alpha\} ]^-$ is an open set. Then 
 $\mathcal{L}_1$ is also Borel because $\mathcal{L}_1  \leq_W \mathcal{L}_2$  and thus  $L$ is also Borel and  the game 
$W(L, L')$ is determined. 

Assume now that $L'$ is not Borel, and consider the Wadge game $W(L, \emptyset + L')$. 

We claim that Player 2 has a w.s. in that 
game which is easily deduced from a w.s. of Player 2 in the Wadge game  $W(\mathcal{L}_1, \mathcal{L}_2)$. 
Consider a play in this latter game where  Player 
1 remains in the closed set $h_1(\Si_1^{\om}) \times \{\alpha\}$:  she writes a beginning of a word in the form 

$$(0.Ax(1).0^2.x(2).0^3.A.x(3) \ldots 0^{2n}.x(2n).0^{2n+1}\ldots ~;~
0.AA.0^2.A.0^3.AA. \ldots AA.0^{2n}.A.0^{2n+1}\ldots )$$

\noi Then player 2 writes a beginning of a word in the form 
$$(0.Ax'(1).0^2.x'(2).0^3.A.x'(3) \ldots 0^{2p}.x'(2p).0^{2p+1}\ldots ~;~
0.AA.0^2.A.0^3.AA. \ldots AA.0^{2p}.A.0^{2p+1}\ldots ) $$
\noi where $p\leq n$. 
Then the strategy for Player 2 in $W(L, \emptyset + L')$ consists to write $x'(1).x'(2) \ldots 
x'(p).$ when Player 1 writes  $x(1).x(2) \ldots x(n).$. If the strategy for Player 2 in $W(\mathcal{L}_1, \mathcal{L}_2)$ was at some step to go out of
the set $h_2(\Si_2^{\om}) \times \{\alpha\}$ then this means that his final word is {\it surely inside} $\mathcal{L}_2$, and that the final word of Player 1 
is also surely inside $\mathcal{L}_1$, because Player 2 wins the play. 
Then Player 2 in the Wadge game $W(L, \emptyset + L')$ can make as he is now in charge of the wholeset and play anything 
(without skipping anymore)  so that  his final $\om$-word is also inside 
$\emptyset + L'$.  So we have proved that Player 2 has a w.s. in  the Wadge game $W(L, \emptyset + L')$  or equivalently that 
$L \leq_W  \emptyset + L'$. But by  Lemma \ref{w1} we know that $L'  \equiv_W \emptyset + L'$  and thus 
$L \leq_W L'$ which means that Player 2 has a  w.s. in  the Wadge game $W(L, L')$. 

\hs \noi {\bf Second case.} Player 1 has a w.s. in the game $W(\mathcal{L}_1, \mathcal{L}_2)$. 

Notice that this implies that 
$\mathcal{L}_2 \leq_W \mathcal{L}_1^-$. Thus if $L$ is Borel then $\mathcal{L}_1$ is Borel, $\mathcal{L}_1^-$ is also Borel, 
and  $\mathcal{L}_2$ is Borel as the inverse image of a Borel set by a continuous function, and thus $L'$ is also Borel, so the Wadge game 
$W(L, L')$ is determined. We assume now that $L$ is not Borel and we consider the Wadge game $W(L, L')$. 
Player 1 has a w.s. in this game which is easily constructed from a w.s. of the same player in the game  $W(\mathcal{L}_1, \mathcal{L}_2)$ as follows. 
For this consider a play in this latter game where Player 2 does not go out of the closed set $h_2(\Si_2^{\om}) \times \{\alpha\}$. 
Then player 2 writes a beginning of a word in the form 
$$(0.Ax'(1).0^2.x'(2).0^3.A.x'(3) \ldots 0^{2p}.x'(2p).0^{2p+1}\ldots ~;~
0.AA.0^2.A.0^3.AA. \ldots AA.0^{2p}.A.0^{2p+1}\ldots )  $$
Player 1, following her w.s. composes 
a beginning of a word in the form 
$$(0.Ax(1).0^2.x(2).0^3.A.x(3) \ldots 0^{2n}.x(2n).0^{2n+1}\ldots ~;~
0.AA.0^2.A.0^3.AA. \ldots AA.0^{2n}.A.0^{2n+1}\ldots )$$
\noi where $p\leq n$. Then the strategy for Player 1 in $W(L,  L')$ consists to write $x(1).x(2) \ldots 
x(n)$ when Player 2 writes  $x'(1).x'(2) \ldots x'(p)$. 

If the strategy for Player 1 in $W(\mathcal{L}_1, \mathcal{L}_2)$ was at some step to go out of
the closed set $h_1(\Si_1^{\om}) \times \{\alpha\}$ then this means that her final word is surely inside $\mathcal{L}_1$, and that the final word of Player 2 
is also surely outside  the set  $\mathcal{L}_2$ (at least if he produces really an infinite word in $\om$ steps). This case is actually not possible because 
Player 2  can always go out of the closed set  $h_2(\Si_2^{\om}) \times \{\alpha\}$ and then his final word is surely in the set $\mathcal{L}_2$.

We have then proved that Player 1  has a w.s. in  the Wadge game $W(L,  L')$.
\ep 

\hs In order to prove our next result we recall that the  following result was proved in \cite{Fin-ICST}. 

\begin{theorem}\label{the5}
 There exists a  $2$-tape B\"uchi automaton $\mathcal{A}$, which can be effectively  constructed, such that the topological complexity of the 
infinitary rational relation  $L(\mathcal{A})$ is not determined by the axiomatic system {\rm ZFC}. Indeed it holds that : 
\begin{enumerate}
\item[(1)] ({\rm ZFC + V=L}). ~~~~~~ The $\om$-language $L(\mathcal{A})$ is an analytic but non-Borel  set. 
\item[(2)] ({\rm ZFC} + $\om_1^{\bf L} < \om_1$).  ~~~~The $\om$-language $L(\mathcal{A})$ is a  ${\bf \Pi}^0_2$-set. 
\end{enumerate}
\end{theorem}

We now state the following new result. 

\begin{theorem}\label{the6}
 Let $\mathcal{B}$ be a B\"uchi automaton accepting the regular   $\om$-language  
$(0^\star.1)^\om \subseteq \{0, 1\}^\om$. Then one can effectively construct a  $2$-tape
B\"uchi automaton $\mathcal{A}$ such that:  
\begin{enumerate}
\item[(1)]  ({\rm ZFC} + $\om_1^{\bf L} < \om_1$).  Player 2 has a winning strategy $F$ in the Wadge game $W(L(\mathcal{A}), L(\mathcal{B}))$. 
But $F$ can not be recursive and not even in the class   $(\Sigma_2^1 \cup \Pi_2^1)$. 
\item[(2)]  ({\rm ZFC} + $\om_1^{\bf L} = \om_1$). The Wadge game $W(L(\mathcal{A}), L(\mathcal{B}))$ 
 is not determined. 
\end{enumerate}
\end{theorem}

\proo It is very similar to the proof of  \cite[Theorem 4.12]{Fin13-JSL}, replacing ``$1$-counter automaton" by 
``$2$-tape B\"uchi automaton" and using the above Theorem \ref{the5} instead of the corresponding result for a real-time $1$-counter automaton 
proved in \cite{Fin-ICST}. In the proof we use in particular  the above Theorem \ref{the5},  the link between Wadge games and 
Wadge reducibility, the ${\bf \Pi}^0_2$-completeness of the regular   $\om$-language  
$(0^\star.1)^\om \subseteq \{0, 1\}^\om$,  the Shoenfield's Absoluteness Theorem, and the notion of  extensions of a model of  {\rm ZFC}. 
 \ep

\hs Notice that
every model of  {\rm ZFC} is either a model of ({\rm ZFC} + $\om_1^{\bf L} < \om_1$) or a model of 
({\rm ZFC} + $\om_1^{\bf L} = \om_1$). Thus  there are no models of   {\rm ZFC}  in which 
Player 1  has a winning strategy in the Wadge game $W(L(\mathcal{A}), L(\mathcal{B}))$. 

\hs Notice also that,  to prove Theorems  \ref{the5} and   \ref{the6},  
we do not need to use any large cardinal axiom  or even the consistency of such an axiom, like the 
axiom of analytic determinacy.

\section{Concluding remarks}
We have proved that the  determinacy of   Gale-Stewart games whose winning sets are accepted by {\it non-deterministic} 
 $2$-tape B\"uchi automata  is  equivalent to the determinacy of (effective) analytic Gale-Stewart games which is known to be a large cardinal 
assumption equivalent to the existence of the real $0^\sharp$. Then we have  proved that the  winning strategies in these games, when they exist, 
 may be very complex, i.e. highly non-effective. Moreover we have proved that, even if we know that some of these games are determined, 
it may be highly undecidable to determine whether Player 1 has a winning strategy. 

On the other hand, we know that the infinitary rational relations accepted by  {\it deterministic} 
 $2$-tape B\"uchi automata are always Borel ${\bf \Delta}_3^0$-sets. Thus this implies that 
Gale-Stewart games whose winning sets are accepted by {\it deterministic} 
 $2$-tape B\"uchi automata are always determined.  It would be interesting to study these games for which the following questions naturally 
arises: can we decide who the winner  is  in such a game?  can we 
compute a winning strategy given by a transducer?

\newcommand{\etalchar}[1]{$^{#1}$}

\end{document}